\documentclass[prb,aps,amssymb,amsmath,reprint,superscriptaddress]{revtex4-1}

\usepackage{lineno,hyperref}

\usepackage{graphicx}
\usepackage[T1]{fontenc}
\usepackage{booktabs}
\usepackage{mathrsfs}
\usepackage{miller}
\usepackage{amsfonts}
\usepackage{xspace}
\usepackage{braket}
\usepackage{siunitx}
\usepackage{lineno}
\usepackage{hyperref} 
\usepackage{multirow}
\usepackage{bigdelim}
\usepackage{tabularx}
\usepackage{extdash}
\usepackage[version=4]{mhchem}
\modulolinenumbers[5]

\DeclareSIUnit\electrons{e\textsuperscript{--}}
\DeclareSIUnit\neutrons{neutrons}
\DeclareSIUnit\ppm{ppm}
\DeclareSIUnit\ppb{ppb}
\DeclareSIUnit\lines{l}
\DeclareSIUnit{\calorie}{cal}

\newcommand{\Nfif}{\ensuremath{^{15}\mathrm{N}}\xspace}

\newcommand{\PtwoCons}{\ensuremath{\mathrm{N}_{3}\mathrm{V}}\xspace}

\newcommand{\NVnb}{\ensuremath{\mathrm{NV}}\xspace}

\newcommand{\NVminus}{\ensuremath{\NVnb^{-}}\xspace}

\newcommand{\NV}{\NVnb\xspace}

\newcommand{\NVN}{\ensuremath{\mathrm{N_{2}V}}\xspace}

\newcommand{\Ns}{\ensuremath{\mathrm{N_{s}}}\xspace}

\newcommand{\Nsneutral}{\ensuremath{\mathrm{N_{s}^{0}}}\xspace}
\newcommand{\NsPlus}{\ensuremath{\mathrm{N_{s}^{+}}}\xspace}

\newcommand{\Sis}{\ensuremath{\mathrm{Si_s}}\xspace}

\newcommand{\SiVneutral}{\ensuremath{\mathrm{SiV}^{0}}\xspace}
\newcommand{\SiVminus}{\ensuremath{\mathrm{SiV}^{-}}\xspace}
\newcommand{\SiVtwominus}{\ensuremath{\mathrm{SiV}^{2-}}\xspace}
\newcommand{\SiV}{\ensuremath{\mathrm{SiV}}\xspace}
\newcommand{\SiVN}{\ensuremath{\mathrm{SiVN}}\xspace}
\newcommand{\SiVNneutral}{\ensuremath{\mathrm{SiVN}^0}\xspace}
\newcommand{\SiVNminus}{\ensuremath{\mathrm{SiVN}^-}\xspace}
\newcommand{\SiVNtwominus}{\ensuremath{\mathrm{SiVN}^{2-}}\xspace}

\newcommand{\SiVHneutral}{\ensuremath{\mathrm{SiVH}^0}\xspace}

\newcommand{\SiVIIHneutral}{\ensuremath{\mathrm{SiV_2H}^0}\xspace}

\newcommand{\SiBneutral}{\ensuremath{\mathrm{SiB}^0}\xspace}

\newcommand{\NVHnb}{\ensuremath{\mathrm{NVH}}\xspace} 
\newcommand{\NVHminus}{\ensuremath{\NVHnb^{-}}\xspace}
\newcommand{\NVHneutral}{\ensuremath{\NVHnb^{0}}\xspace}
\newcommand{\NIIVHnb}{\ensuremath{\mathrm{N_2VH}}\xspace} 
\newcommand{\NIIVHneutral}{\ensuremath{\NIIVHnb^{0}}\xspace}
\newcommand{\NIIIVHnb}{\ensuremath{\mathrm{N_3VH}}\xspace} 
\newcommand{\NIIIVHneutral}{\ensuremath{\NIIIVHnb^{0}}\xspace}

\newcommand{\MonoclinicI}{\ensuremath{\mathrm{C}_{\mathrm{1h}}}\xspace}

\mathchardef\mhyphen="2D

\newcommand{\asgrown}{Sample~A\xspace}

\newcommand{\treatedth}{Sample~D\xspace} 
 
\newcommand{\treatedtfh}{Sample~F\xspace} 
\newcommand{\nfourtreated}{Sample~G\xspace} 

\newcommand{\chspin}[2]{\ensuremath{{}^{#1}\mathrm{X^{#2}}}\xspace}
\newcommand{\Xminus}{\chspin{}{-}}
\newcommand{\Xneutral}{\chspin{}{0}}
\newcommand{\Xplus}{\chspin{}{+}}

\setcitestyle{numbers,square}

\begin{document}

\title{Doubly-charged silicon vacancy center, photochromism, and Si-N complexes in co-doped diamond}

\author{B~G~Breeze}
\affiliation{Department of Physics, University of Warwick, Coventry, CV4 7AL, UK}
\author{C~J~Meara}
\affiliation{School of Engineering, Newcastle University, Newcastle upon Tyne NE1 7RU, UK}
\affiliation{EPSRC Centre for Doctoral Training in Diamond Science and Technology, UK}
\author{X~X~Wu}
\author{C~P~Michaels}
\author{R~Gupta}
\affiliation{Department of Physics, University of Warwick, Coventry, CV4 7AL, UK}
\author{P~L~Diggle}
\author{M~W~Dale}
\author{B~L~Cann}
\affiliation{De Beers Group Technology, Maidenhead, Berkshire, SL6 6JW, UK}
\author{T~Ardon}
\author{U~F~S~D'Haenens-Johansson}
\affiliation{Gemological Institute of America, 50 W 47th Street, New York, New York 10036, USA}
\author{I~Friel}
\affiliation{Element Six, Global Innovation Centre, Fermi Avenue, Didcot OX11 0QR, UK}
\author{M~J~Rayson}
\author{P~R~Briddon}
\author{J~P~Goss}
\affiliation{School of Engineering, Newcastle University, Newcastle upon Tyne NE1 7RU, UK}
\author{M~E~Newton}
\affiliation{Department of Physics, University of Warwick, Coventry, CV4 7AL, UK}
\affiliation{EPSRC Centre for Doctoral Training in Diamond Science and Technology, UK}
\author{B~L~Green}
\email[Corresponding author: ]{b.green@warwick.ac.uk}
\affiliation{Department of Physics, University of Warwick, Coventry, CV4 7AL, UK}

\begin{abstract}
We report the first experimental observation of a doubly-charged defect in diamond, \SiVtwominus{}, in silicon and nitrogen co-doped samples. We measure spectroscopic signatures we attribute to substitutional silicon in diamond, and identify a silicon-vacancy complex decorated with a nearest-neighbor nitrogen, \SiVN, supported by theoretical calculations. Samples containing silicon and nitrogen are shown to be heavily photochromic, with the dominant visible changes due to the loss of $\mathrm{SiV^{0/-}}$ and gain in the optically-inactive \SiVtwominus{}. 
\end{abstract}
\maketitle

\section{Introduction}
Diamond, as with other wide-band-gap semiconductors, has recently attracted attention as a host for optically-active point defects with potential applications in quantum communication \cite{Atature2018}, nanophotonics \cite{Aharonovich2014,Schroder2016}, and quantum information processing (QIP) \cite{Awschalom2018}. In addition to the well-known nitrogen-vacancy (NV, where V denotes a vacancy henceforth) \cite{Doherty2013}, the group-IV-vacancy centres (SiV \cite{Evans2018,Rose2018,Green2017c}, SnV \cite{Iwasaki2017}, and PbV \cite{Trusheim2019}) have recently emerged as potential candidates in QIP applications. 

Unlike bulk nitrogen-doped diamond, where a significant effort stretching over decades has identified many nitrogen-related point defects \cite{Zaitsev2001,Dischler2012}, relatively little experimental study has been performed on high-quality single crystal diamond which is bulk-doped with silicon. The only definitive assignments of optical centers to silicon are the well-known \SiVminus{} \cite{Goss1996,Rogers2014a,Hepp2014a} and \SiVneutral{} \cite{DHaenens-Johansson2011,Green2017c,Green2019}. Additionally, electron paramagnetic resonance (EPR) studies have identified \SiVHneutral \cite{Iakoubovskii2003a,Edmonds2008a} and \SiVIIHneutral \cite{DHaenens-Johansson2010}, while a tentative assignment has been made to \SiBneutral \cite{Nadolinny2016}. Density functional theory (DFT) studies of silicon-related point defects indicate that isolated substitutional silicon, \Sis, is stable though aggregates are energetically unfavorable \cite{Goss2007}. Some silicon-related multi-vacancy, multi-hydrogen, nitrogen-related complexes are theoretically stable \cite{Goss2007,Thiering2015} but most have yet to be identified experimentally.

In this work, we have studied silicon and nitrogen co-doped single-crystal synthetic diamond from as-grown to a treatment temperature of \SI{2400}{\celsius} using a combination of optical absorption spectroscopies and EPR. We identify the neutrally-charged silicon-vacancy-nitrogen complex \SiVNneutral{} through combined experimental measurements and theoretical modeling, and tentatively assign an infrared absorption mode at \SI{1338}{\per\centi\meter} to substitutional silicon.

\subsection{Charge transfer}
\label{subsec:charge_transfer}
It is well established that as an insulator, defects in diamond may exist in more than one charge state in the same crystal simultaneously. For charge states which are dominated by the charge dynamics of nitrogen donors (the dominant impurity in the majority of synthetic diamond), a ``charge transfer'' protocol has been established to drive between the two extremal states \cite{DHaenens-Johansson2011,Khan2009,Khan2013}. Above-band-gap UV excitation ({$\lambda\SI{<225}{\nano\meter}$}) of a sample typically maximizes the concentration of nitrogen donors, \Nsneutral{}, which in turn tends to favor the neutral charge state of other defects. Heating the sample at \SI{550}{\celsius} in the absence of light enables thermal excitation of electrons/holes, reversing the process and yielding \NsPlus{} while typically maximizing the concentrations of negatively-charged versions of defects present. This has been previously demonstrated in several defects including \SiV \cite{DHaenens-Johansson2011}, \NVnb{} \cite{Maki2011}, \NVHnb \cite{Khan2009}, \NVN{} \cite{Dale2015}, and \PtwoCons \cite{Green2017}.

This charge instability can be a great advantage when studying fundamental defect properties as it enables multiple charge states of the same defect to be studied in the same crystal. In a single-electron charge transfer process (e.g. \Xneutral to \Xminus rather than \Xplus to \Xminus for a given defect X), we expect at least one of the charge states to be EPR-active. EPR is capable of absolutely quantifying the concentration of a defect and is therefore fundamentally the source of the optical absorption cross-section values for most defects in diamond \cite{Davies1999}. Where charge transfer is present, the loss (gain) of the EPR-active charge state can be equated to the gain (loss) in the other, allowing an optical absorption cross-section to be extracted for the non-EPR-active state. The major assumption made in the above procedure is that there are only two charge states of the defect accessible through the charge transfer protocol, and therefore the loss of one must equal the gain in the other. If this is not the case, and some charge population is lost to a third charge state, then an incorrect optical absorption cross-section for the non-EPR-accessible charge state will be extracted.

\section{Experimental detail}
\subsection{Method and samples}
Seven samples were grown in a microwave-plasma chemical vapor deposition (CVD) reactor: Samples~A -- F were grown simultaneously and doped with natural abundance silicon (via the addition of silane to the growth gases) and \SI{100}{\percent} \Nfif{}-enriched nitrogen; \nfourtreated was doped with natural abundance silicon and nitrogen (Table~\ref{tab:sample_summary}). Growth substrates (all \hkl{001}-oriented) and non-diamond material were removed from all samples post-growth to leave free-standing plates. Samples~B -- F were each subsequently annealed under stabilizing pressure for \SI{1}{\hour} at {1600,~1800,~...,~\SI{2400}{\celsius}}, respectively: each sample was annealed only once (i.e. \treatedth{} was annealed at \SI{2000}{\celsius} only). \nfourtreated was annealed at \SI{1800}{\celsius} for \SI{100}{\hour} under stabilizing pressure. All samples were polished post-anneal to remove any etched or graphitic material and provide parallel, low-roughness faces for optical measurements. Each sample was approximately $3\times3\times\SI{1.6}{\milli\meter}$.

As a consequence of the charge transfer effect ($\S\ref{subsec:charge_transfer}$), the annealing behavior of a given defect can be confused with its charge transfer properties if care is not taken to initialize the crystal to a known state before each measurement. We therefore perform all measurements immediately following either UV exposure (the ``UV state'') or heating in the dark at \SI{550}{\celsius} for \SI{20}{\minute} (the ``heated state''): the sample is kept in the dark between treatment and measurement. For the UV state, samples were exposed using the xenon arc lamp of a DiamondView instrument for \SI{6}{\minute} per face; heating was performed under a dry nitrogen atmosphere in a tube furnace (Elite Thermal Systems Ltd.) for \SI{20}{\minute} at \SI{550}{\celsius}. EPR measurements were performed at X-band using a Bruker EMX-E spectrometer with \SI{90}{\decibel} attenuator to avoid microwave power saturation, and ER4122SHQ resonator. EPR measurements were quantified by comparison to a standard reference sample containing \SI{270}{\ppm} \Nsneutral{} and were performed below microwave power saturation. Ultraviolet-visible (UV-vis) and IR absorption measurements were performed in PerkinElmer Lambda~1050 and Spectrum~GX spectrometers, respectively. 

\begin{table}[tb]
\centering
\setlength{\tabcolsep}{4pt}
\caption{Summary of the samples employed in this study. All post-growth anneals were performed under stabilizing pressure and for \SI{1}{\hour}, except for \nfourtreated which was annealed for \SI{100}{\hour}. Dopants without explicit isotopes are natural abundance.}
\label{tab:sample_summary}
\begin{tabular}{
	c
	p{0.3cm}
	c
	c
	r
	}
	\toprule
	
	Sample 	&& 	Dopants		& 	Growth run 	& Annealing temp (\si{\celsius})\\
	\midrule
	A 	   	&&  \Nfif, Si 	& 	1			& As-grown\\
	B 		&& 	\Nfif, Si 	& 	1			& 1600\\
	C 		&& 	\Nfif, Si 	& 	1			& 1800\\
	D 		&& 	\Nfif, Si 	& 	1			& 2000\\
	E 		&& 	\Nfif, Si 	& 	1			& 2200\\
	F 		&& 	\Nfif, Si 	& 	1			& 2400\\
	G		&& 	N, Si 	& 	2			& 1800\\
	\bottomrule
\end{tabular}
\end{table}

\subsection{Computational method}
Density functional theory within the supercell approach was employed using the AIMPRO software package\,\cite{rayson2009highly}.  We have used a generalized gradient approximation (GGA)\,\cite{perdew1996generalized} for the exchange and correlation and the pseudo-potential approximation\,\cite{hartwigsen1998relativistic} to remove the core electrons from explicit determination.
Kohn-Sham functions were expanded in a basis of atom-centered Gaussian functions\,\cite{Goss2007} using four d-type functions resulting in 40 functions per atom. 
The charge density was Fourier transformed using plane waves with a cutoff of 300\,Ha, which results in total energy convergence to \SI{1e-5}{\electronvolt} with respect to this parameter. 
The Brillouin zone was sampled using the Monkhorst-Pack scheme \cite{monkhorst1976special}: the maximum reciprocal volume per sampling point was \num{0.01}. 

Using this approach, the lattice constant of diamond agrees with experiment (\SI{3.57}{\angstrom}\,\cite{riley1944lattice}) to within \SI{1}{\percent}.
All defect structures were modeled using simple-cubic supercells based upon the 8-atom conventional unit cell, with lattice spacing of $4a_0$ containing 512 atoms.

Donor and acceptor levels were found using the formation energy method\,\cite{goss2004,chemp}, with the formation energy ($E^f$) for a certain charge state, $q$, obtained using
\begin{equation}
\label{eq:form}
E^f(X,q) = E_{\text{tot}}(X,q) - \sum \mu_i + q(E^X_V + \mu_e) + \chi(X,q).
\end{equation}
Here $E_{\text{tot}}$ is the total energy of a defect structure, $E^X_V$ is taken as the valence-band maximum, $\mu_e$ is the electron chemical potential and $\chi$ is a correction term for periodic charge in the supercell \cite{shim2005density}.
$\chi$ comprised of a background electrostatic correction of \si{\milli\electronvolt} order and the Madelung term for the $4a_0$ supercell calculated at \num{0.26}$q^2$~\si{\electronvolt}. 
Binding energies were calculated using formation energies \cite{goss2008dissociation,goss2004}, as the energy released in the formation of the complex from the component parts.
Hyperfine tensor principal values and directions were determined as described previously \cite{Goss2007,shaw2005importance}. 

\section{Results}
\subsection{Annealing}
\label{subsec:annealing}
Initially we consider the annealing study performed on the samples which were grown simultaneously (samples A--F). Each sample was initialized into the UV state and measured by IR and EPR to quantify the defects present --- see Table~\ref{tab:quantification_methods} for details on quantification method for each defect. 

\begin{figure}[tbhp]
\centering
\includegraphics[width=\columnwidth]{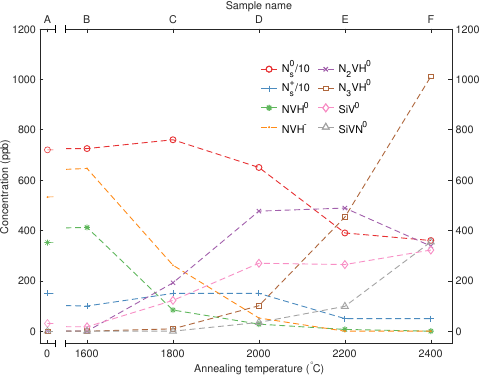}
\caption{Point defect concentrations measured by EPR and IR in samples grown simultaneously and subsequently annealed for \SI{1}{\hour} at high temperature under stabilizing pressure (see Table~\ref{tab:sample_summary}). All measurements taken with the sample in the UV-treated charge state (see text for details). Dashed lines are guides to the eye.}
\label{fig:annealing}
\end{figure}

\asgrown is dominated by nitrogen-related complexes, with the most abundant identified defects being \Nsneutral{}$^{/+}$ and \NVHneutral{}$^{/-}$ [Fig.~\ref{fig:annealing}]. The only identified silicon-related centres are \SiVneutral{}$^{/-}$ which are present at approximately \SI{100}{\ppb} combined. Any concentrations of \SiVHneutral{} \cite{Edmonds2008a} and \SiVIIHneutral{} \cite{DHaenens-Johansson2010} are below EPR detection limits ($\SI{\approx1}{\ppb}$). \asgrown is visually brown but heavily photo/thermochromic, varying from deep brown to brown-pink in the UV and heated states, respectively. 

\begin{table}[tb]
\centering
\caption{Details and references for techniques and absorption cross-section coefficients employed in the quantification of defects at each annealing stage.}
\begin{tabular}{lp{0.3cm}lp{0.3cm}Sc}
\toprule
Defect && Technique && \multicolumn{1}{l}{Note} & \multicolumn{1}{c}{Ref.} \\
\midrule
\Nsneutral{} 	&& IR 		&& 	\SI{1344}{\per\centi\meter} & \cite{Woods1990} \\
\NsPlus{} 		&& IR 		&& 	\SI{1332}{\per\centi\meter} & \cite{Lawson1998} \\

\NVHneutral{} 	&& IR 		&& 	\SI{3123}{\per\centi\meter} & \cite{Liggins2010b} \\
\NVHminus{} 	&& EPR 		&& 								& \cite{Glover2003a} \\
\NIIVHneutral{} && EPR 		&& 								& \cite{Hartland2014} \\
\NIIIVHneutral{}&& IR 		&&  \SI{3107}{\per\centi\meter}	& \cite{Hartland2014} \\
\SiVneutral{} 	&& EPR 		&&								& \cite{DHaenens-Johansson2011} \\
\SiVminus{} 	&& UV-vis 	&&  \SI{737}{\nano\meter} 		& \cite{DHaenens-Johansson2011} \\
\SiVNneutral{} 	&& EPR 		&&  \multicolumn{1}{c}{\textit{this work}} \\
\bottomrule
\end{tabular}
\label{tab:quantification_methods}
\end{table}

Analagous to the well-known aggregation of nitrogen-vacancy centers in diamond ($\mathrm{N}_n\mathrm{V}$, where $n=\numrange{1}{4}$) \cite{Evans1982}, we observe the aggregation of $\mathrm{N}_n\mathrm{VH}$ as the annealing temperature increases. A decrease in \NVHnb{} at $\geq\SI{1800}{\celsius}$ and above is accompanied by an increase in \NIIVHneutral{}, which in turn decreases at \SI{2400}{\celsius} with a corresponding rise in \NIIIVHneutral{} [Fig~\ref{fig:annealing}]. 

\begin{figure}[tb]
\centering
\includegraphics[width=\columnwidth]{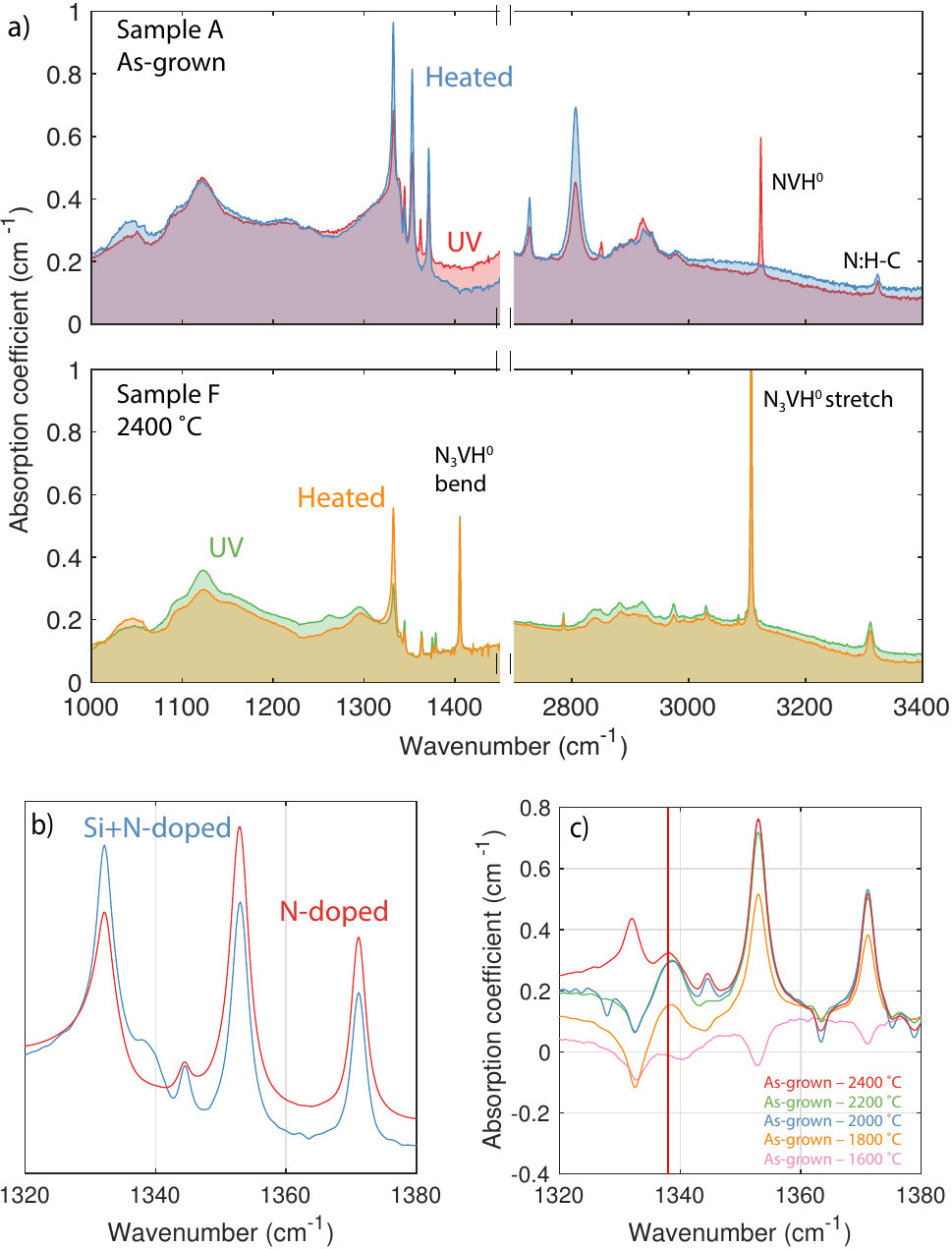}
\caption{a) IR absorption spectra of \asgrown (top) and \treatedtfh (bottom) in the two extremal charge states. The two regions give the defect-induced one-phonon absorption (left) and C-H stretch region (right): the intrinsic multiphonon absorption has been subtracted. b) The one-phonon of two as-grown samples grown under similar CVD conditions: both are nitrogen-doped but silicon was added to the growth gasses of one. The primary difference is the feature at approx.\ \SI{1340}{\per\centi\meter} in the silicon-containing sample, which is tentatively assigned to substitutional silicon \cite{Goss2007}. The remaining peaks are reported in studies of solely nitrogen-doped CVD material \cite{Khan2009}. c) Difference spectra between \asgrown (as-grown) and samples treated at the given temperatures. The change in the \SI{1338}{\per\centi\meter} mode is highlighted.}
\label{fig:ir_comparison}
\end{figure}

The majority of the sharp IR one-phonon and C-H stretch absorption peaks observed in \asgrown [Fig.~\ref{fig:ir_comparison}a)] have been previously observed in high-nitrogen, high-hydrogen brown diamond from several sources \cite{Wang2007,Khan2010}, and their photochromic behavior reported \cite{Khan2010}. The point defect origin of these peaks has not been identified, but they do not appear to require silicon. However, the small shoulder at \SI{1338}{\per\centi\meter} is not present in previous reports of high-nitrogen material. The peak itself is not photochromic, and its frequency does not depend on nitrogen isotope. Samples grown under similar conditions but without the addition of silicon to the growth gasses produce similar one-phonon spectra except for the absence of the \SI{1338}{\per\centi\meter} mode [Fig.~\ref{fig:ir_comparison}(b)]. Previous DFT calculations predict a mode originating at the carbon atoms surrounding substitutional silicon at \SI{1333}{\per\centi\meter} \cite{Goss2007}: in conjunction with studies of silicon-doped HPHT-grown samples \cite{Palyanov2017,Palyanov2015b}, we tentatively assign the \SI{1338}{\per\centi\meter} peak to substitutional silicon. Difference spectra reveal essentially no change between the as-grown and \SI{1600}{\celsius} samples, with subsequent anneals reducing the strength of the \SI{1338}{\per\centi\meter} mode [Fig.~\ref{fig:ir_comparison}(c)]--- this is consistent with the increase in observed Si-related defects from \asgrown to \treatedtfh [Fig~\ref{fig:annealing}].

The concentration of \SiVneutral{}$^{/-}$ increases by over an order of magnitude from \asgrown to \treatedtfh. We conclude that the majority of the silicon was originally incorporated in other forms (assumedly substitutionally) during growth, with the subsequent production of \SiV{} proceeding by vacancy capture during the HPHT treatment, analagous to the production of the $\mathrm{N}_n\mathrm{VH}$ defects. EPR measurements of samples annealed at \SI{2000}{\celsius} and higher reveal the presence of a previously-unidentified silicon-containing defect. We identify this defect as a silicon-vacancy center decorated with a nitrogen atom (\SiVNneutral{}): the defect is discussed further in \S\ref{sec:sivn}. The concentration of \SiVneutral and \SiVNneutral{} measured in \treatedtfh [Fig.~\ref{fig:annealing}] indicates that at least \SI{1}{\ppm} of silicon must have been incorporated during growth. 

Between samples A and F, approximately \SI{4.5}{\ppm} of substitutional nitrogen has been lost in addition to \SI{1}{\ppm} of \NVHneutral{}$^{/-}$, and is accompanied by the production of approximately \num{0.3} and \SI{1.0}{\ppm} of \NIIVHneutral{} and \NIIIVHneutral{}, respectively. Together with \SiVNneutral{} this corresponds to a total of \SI{4.0}{\ppm} nitrogen, accounting for the majority of the lost \Nsneutral{}$^{/+}$ and \NVHneutral{}$^{-}$. However, a significant concentration of nitrogen-related defects remain unidentified. As \NIIIVHnb{} contains three nitrogen atoms, a small error in its oscillator strength would have a dramatic effect on our ability to quantify total nitrogen in the high-temperature annealed samples. 

\subsection{Photochromism and evidence for \texorpdfstring{$\mathbf{SiV^{2-}}$}{SiV2-}}
In previous studies of nitrogen-doped brown CVD diamond, samples which were annealed above \SI{1600}{\celsius} became less brown, with higher temperatures corresponding to a greater reduction of brown color \cite{Martineau2004,Khan2009,Khan2013}. The present samples display the same behavior, with samples~E and F (\num{2200} and \SI{2400}{\celsius}, respectively) appearing near-colorless by eye in the heated state. Contrary to previous studies, the present samples treated at \SI{>2000}{\celsius} remain heavily photochromic, varying from a deep grey-blue to near-colorless in the UV and heated states, respectively. 

\begin{figure}[tb]
\centering
\includegraphics[width=\columnwidth]{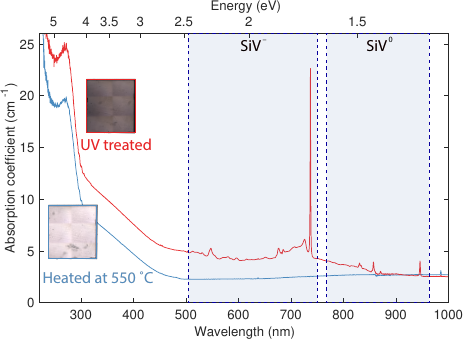}
\caption{UV-Vis absorption spectra of \treatedth (annealed at \SI{2000}{\celsius}) measured at \SI{80}{\kelvin} in two extremal charge states --- see text for details. Strong \SiVminus{} and \SiVneutral{} spectra are recorded in the UV-treated charge state, and are undetectable in the heated charge state. The charge transfer processes are reversible i.e. no net \SiV{} is created or destroyed in during the treatments: the dramatic loss of \SiVminus{} and \SiVneutral{} between extremal states is therefore strong evidence for the existence of a third charge state of \SiV{}, which we identify as \SiVtwominus{}. Note that a broad \SiVneutral{} absorption continues beneath the \SiVminus{} absorption, but the structure between \SIrange{500}{740}{\nano\meter} belongs to \SiVminus{}. Inset: representative transmission images of the sample in the two extremal charge states.}
\label{fig:UV-vis-ch}
\end{figure}

UV-Vis measurements of \treatedth in the UV state show strong absorption from both \SiVminus{} (\SI{737}{\nano\meter}) and \SiVneutral{} (\SI{946}{\nano\meter}) [Fig.~\ref{fig:UV-vis-ch}]. The spectrum of the former reveals the optical structure associated with the second excited state of \SiVminus{} \cite{Rogers2014a,Haußler2017} which has been reported previously in photoluminescence excitation \cite{Ekimov2017a,Ekimov2018}. Comparison of the absorption spectra in the UV and heated states confirms that the photo/thermochromism is dominated by dramatic changes in the concentration of \SiVminus and \SiVneutral{} [Fig.~\ref{fig:UV-vis-ch}]: this is the case for samples A--F. The visible photochromic color change in the present samples is much more extreme than the color change reported in nitrogen-doped CVD samples \cite{Khan2009} due to the incredibly broad absorption band of \SiVneutral{} compared to the relatively broad and weak absorption bands associated with \NVHnb{} in purely nitrogen-doped material [Fig~\ref{fig:UV-vis-ch}]. 

The processes employed during the charge transfer procedure are not capable of destroying or creating \SiV and we conclude that we are efficiently driving to a third charge state of \SiV. The photochromic behavior of the samples is consistent with driving to a negatively- (rather than positively-) charged state. DFT studies of \SiV predict that \SiVtwominus is a stable and electronically saturated system with no internal optical transitions or accessible spin levels and is thus difficult to spectroscopically observe \cite{Gali2013}. We therefore infer the presence of \SiVtwominus{} by the absence of \SiVminus{} and \SiVneutral{} in the heated charge state.

\subsection{Discussion}
Doubly-charged defects are well-characterized in other group-IV semiconductors (e.g. Si~\cite{Watkins2000,Newman1982} and Ge~\cite{Weber2013,Weber2015}) but have not been previously reported in diamond despite several theoretical predictions \cite{Breuer1995,Mainwood1997,Thiering2015}. Generally, this is a result of the paucity of very shallow donors and acceptor states in diamond, which are required to stabilize the chemical potential suitably for these doubly-charged states (in the absence of negative-U effects \cite{Watkins1980}). However, in \SiV the (2$-$/$-$) transition is relatively deep (approximately mid-gap \cite{Gali2013}), yielding a stable charge state even for deep nitrogen donors (at approximately $E_C-\SI{1.7}{\electronvolt}$ \cite{Farrer1969a}). 

Despite its lack of internal transitions, we still expect transitions from the \SiVtwominus{} ground state to the conduction band, which are theoretically predicted at \SI{\approx 4}{\electronvolt} \cite{Gali2013}. As a defect-to-band transition, this will manifest as spectrally broad rather than a sharp transition. There is a small change in the absorption gradient \SI{<250}{\nano\meter} (\SI{<5}{\electronvolt}) between the UV and heated states, but any absorption in this region is dominated by \Nsneutral{} absorption [Fig.~\ref{fig:UV-vis-ch}] \cite{Jones2009b} and hence difficult to isolate. 

In \treatedth we measure the UV state concentrations of \SiVminus{} and \SiVneutral{} as \SI{110}{\ppb} and \SI{380}{\ppb}, respectively [Fig~\ref{fig:UV-vis-ch}], using the conversion factors given in \cite{DHaenens-Johansson2011}: in the heated state the concentration of both charge states is below \SI{1}{\ppb} and therefore all \SiV defects are in the $2-$ charge state, requiring \SI{870}{\ppb} of donor charges between the two states. The corresponding loss in \Nsneutral{} from UV to heated states is \SI{2.3}{\ppm}, more than accounting for the \SiV-related charge effects. This relationship is true at all annealing temperatures. As a result, the changes in donor concentrations cannot be attributed solely to \SiV and it is therefore difficult to quantify the latent \SiVtwominus{} concentration in the UV state. Upper limits can be estimated based on the assumption that the only donor is \Nsneutral{}; however, this is known not to be the case in these samples (e.g. \NVHminus{}, other photochromic peaks in [Fig~\ref{fig:ir_comparison}]). 

The extremal charge states are unstable at room temperature in all of the present samples. Time-lapse absorption measurements of \SiVminus{}, performed in the absence of ambient light, show that after UV treatment of \treatedtfh the concentration of \SiVminus{} increases by approximately \SI{70}{\percent} over \SI{9}{\hour}. Ambient light increases the rate of this change, and significant color changes are visible after \SI{2}{\hour} in ambient. The changes cannot be described by a simple coupled model with constant leakage rates from $\SiVneutral{}\rightarrow\SiVminus{}$ and $\SiVminus{}\rightarrow\SiVtwominus{}$. Instead, the increase is well-described by a hyperbolic function, as expected by multiple overlapping thermal processes. This is consistent with the present material containing multiple thermally-activated donors/acceptors at room temperature. 

The existence of \SiVtwominus{} casts doubt on the optical absorption cross-section for \SiVminus given in \cite{DHaenens-Johansson2011}. The cross-section for \SiVneutral{} was calibrated by directly measuring its concentration by EPR and equating it to the absorption strength measured by UV-Vis. However, the cross-section for \SiVminus{} was calibrated via charge transfer between \SiVneutral{} and \SiVminus{} using the protocol given in \S\ref{subsec:charge_transfer}: the loss of the former was equated to the gain in the latter. The assumption was that only two charge states were involved in the process: any loss or gain of population to or from \SiVtwominus{} was unaccounted for, and would result in a modified absorption cross-section than the one given in \cite{DHaenens-Johansson2011}. The concentrations of \SiVminus{} given by the cross-section are within approximately a factor of two of the expected concentration based on charge balance arguments. However, the high concentration of \SiVtwominus{} in these samples makes a more precise statement impossible at this time. A future study based on intrinsic or even p-type material should bias between \SiVminus{} and \SiVneutral{}, allowing both present charge states to be quantified simultaneously and reliably. We note that even with the present uncertainty, our results remain incompatible with the \SI{1e-13}{\milli\electronvolt\per\centi\meter\squared} value derived from first-principles calculations \cite{Kern2017}.

\section{The silicon-vacancy-nitrogen defect}
\label{sec:sivn}
\subsection{Defect identification}
\begin{figure}[t]
	\centering
	\includegraphics[width=\columnwidth]{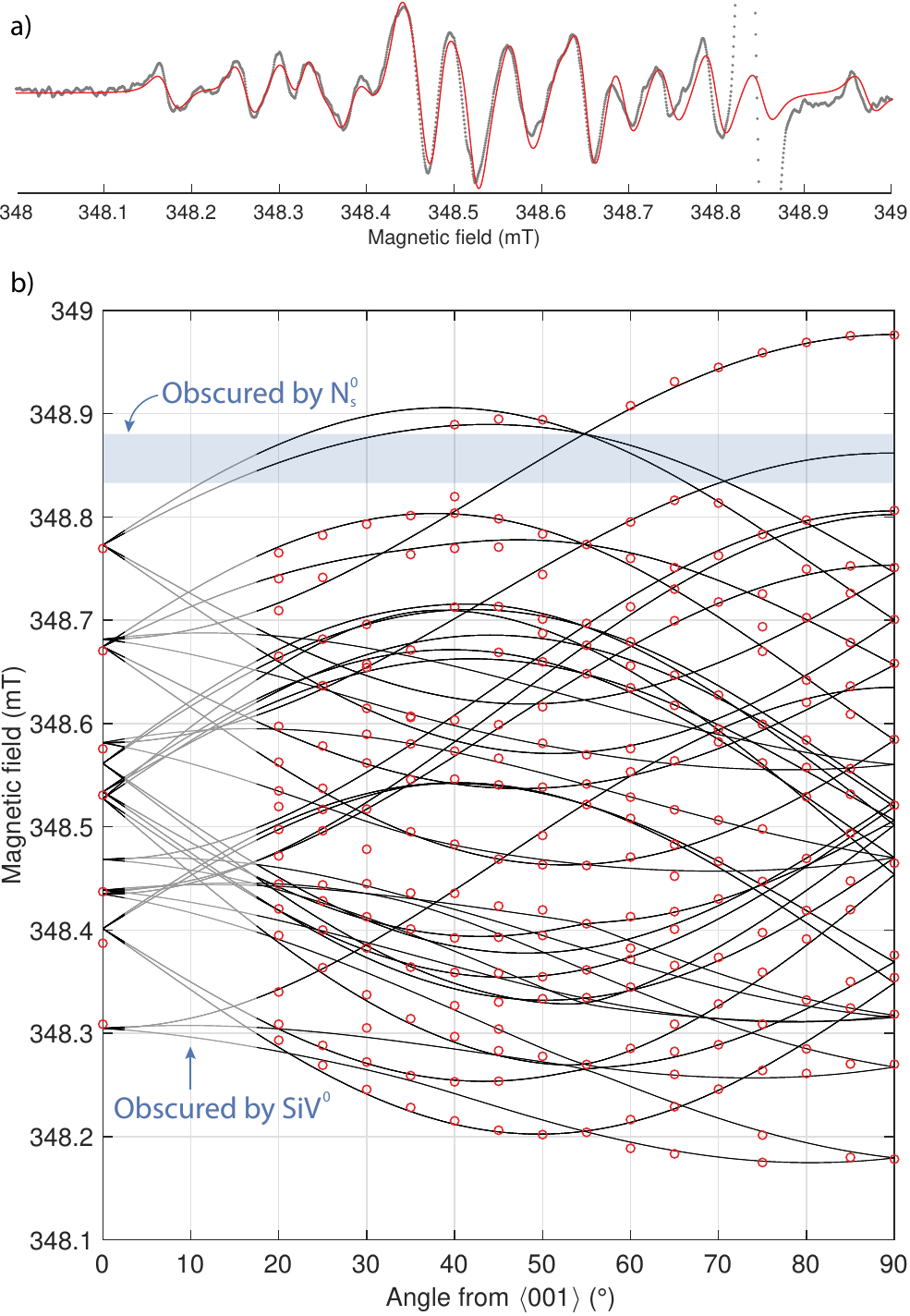}
	\caption{a) EPR spectrum of \nfourtreated with $B\|\hkl<110>$. Experiment in black; simulation in red. b) Angular variation (in a \hkl{110} plane) of measured EPR transition fields (circles) overlaid with a simulation produced using the spin Hamiltonian parameters given in Table~\ref{tab:spin_hamiltonian_parameters}. To improve clarity, only transitions with a theoretical intensity \SI{>=30}{\percent} of the most intense transition are shown.}
	\label{fig:110_and_roadmap}
\end{figure}

EPR measurements of samples D--F reveal a previously-unidentified multi-line ${S=1/2}$ spectrum at approximately $g=\num{2.004}$. Initial analysis indicated a defect which possessed a \SI{100}{\percent} ${I=1/2}$ nucleus with a small hyperfine interaction. As these samples are \Nfif-enriched, the nucleus involved could either be \ce{^15N} or \ce{^1H}. An additional sample, \nfourtreated, which was grown under similar conditions to Samples~A--F but with natural abundance nitrogen rather than \Nfif{}-enriched gasses and subsequently annealed at \SI{1800}{\celsius} for \SI{100}{\hour}, was studied to identify the nucleus involved. We again observe a previously-unidentified multi-line ${S=1/2}$ spectrum at approximately $g=\num{2.004}$, with more transitions than in the \ce{^15N}-doped samples [Fig~\ref{fig:110_and_roadmap}(a)]. In conjunction with the angular variation of the spectrum [Fig~\ref{fig:110_and_roadmap}(b)], the spectrum was identified as belonging to a defect possessing a nuclear spin of \SI{100}{\percent} $I=1$, and a non-zero quadrupole interaction. Due to the isotopic abundances we identify this nucleus as a single nitrogen atom, eliminating hydrogen as a possibility. The defect possesses monoclinic \MonoclinicI{} symmetry and a small hyperfine interaction with the nitrogen [Table~\ref{tab:spin_hamiltonian_parameters}], indicating a low-symmetry defect with essentially zero unpaired electron spin density on the nitrogen nucleus \cite{Green2015}.

A large number of purely nitrogen-related defects have been identified by EPR in diamond, including \Nsneutral{} \cite{Smith1959a}, \NVminus{} \cite{Loubser1977,Loubser1978}, interstitial nitrogen \cite{Felton2009}, and even substitutional nitrogen pairs \cite{Nadolinny1999a}. It is thus unlikely that a new defect which involves only nitrogen would be identified in material which is novel due to its simultaneously high concentration of nitrogen and silicon. Therefore, we hypothesise that this defect must also contain silicon, whose \SI{95}{\percent} natural abundance of $^{28}\mathrm{Si}$ ($I=0$) makes it difficult to identify without a high defect concentration. 

\begin{figure}[t]
	\centering
	\includegraphics[width=\columnwidth]{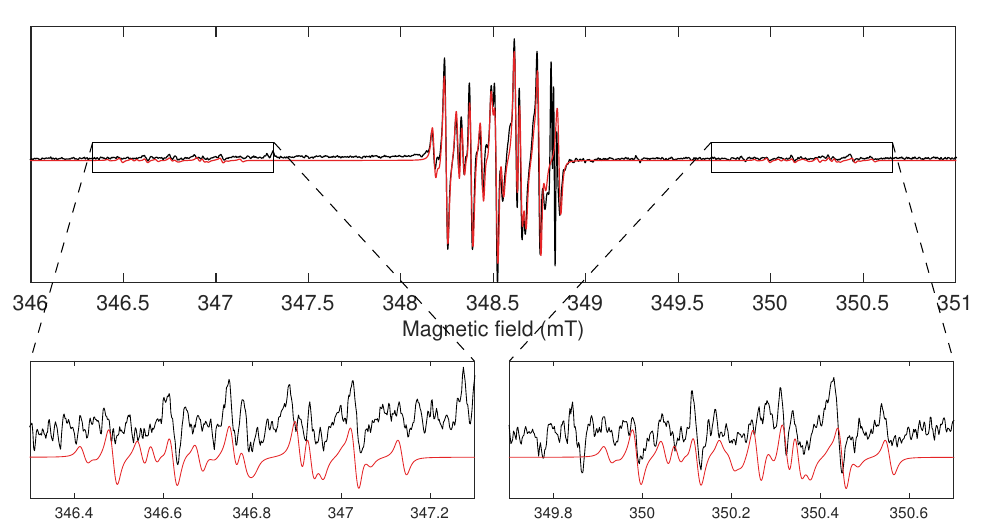}
	\caption{EPR spectrum of \nfourtreated with $B\|\hkl<111>$. Experimental data in black; simulation in red. Additional panels show the $^{29}\mathrm{Si}$ hyperfines on each side of the primary spectrum. As expected, their intensity is \SI{5}{\percent} of the primary spectrum. Simulation generated by EasySpin \cite{Stoll2006} using the spin Hamiltonian parameters given in Table~\ref{tab:spin_hamiltonian_parameters}.}
	\label{fig:si29_epr}
\end{figure}

\begin{figure}[t]
\centering
\includegraphics[width=\columnwidth]{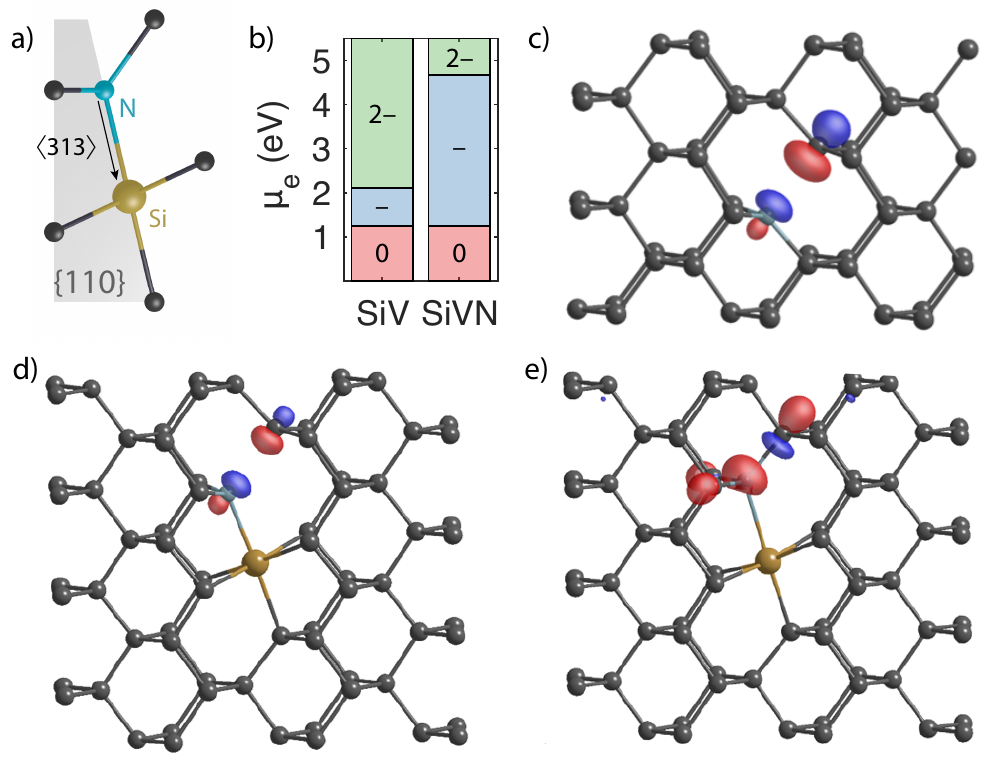}
\caption{a) Schematic of the \SiVN{}$^{0/-}$ defect, highlighting the defect's \hkl<110> mirror plane. b) Calculated formation energies at varying chemical potentials, $\mu_e$, for \SiV and \SiVN with reference to the intrinsic diamond valence band maximum. Transition levels include charge density offset and Madelung corrections. The calculated conduction band minimum was at \SI{4.27}{\electronvolt}. c--e) Electron density on bandgap states for c) the donor state in \Nsneutral{}; d) the state with an unpaired electron for \SiVNtwominus{}; e) the same state for \SiVNneutral{}. Comparison of d) and e) highlights the additional electron in an N-C antibonding orbital in \SiVNtwominus{}. Isosurfaces depict a surface of constant spin density: a common spin density threshold was chosen for d) and e) to allow comparison in the same structure; a higher threshold was chosen for c) due to the highly localized nature of the \Nsneutral{} donor state.}
\label{fig:orbitals}
\end{figure}

A previous DFT study into silicon-containing defects in diamond identified SiVN as a simple and stable defect candidate in high-nitrogen high-silicon diamond \cite{Goss2007}. DFT calculations of the hyperfine parameters of the silicon and nitrogen in \SiVNneutral (improving on previously-reported values \cite{Goss2007}) were used as a guide for experimental parameters [Table~\ref{tab:spin_hamiltonian_parameters}]. To confirm the presence of silicon in the defect, long-term scans designed to increase the signal-to-noise enough to easily identify any $^{29}\mathrm{Si}$-related spectrum (approximately \SI{5}{\percent} of the natural abundance of $^{28}\mathrm{Si}$) were performed. These scans measured approximate replicas of the primary spectrum split by a nucleus of $I=1/2$, approximately \SI{5}{\percent} abundant, which we identify as $^{29}\mathrm{Si}$ [Fig~\ref{fig:si29_epr}]. The hyperfine interaction strengths $A_{1,2,3}=\num{98.24}$, $\num{98.13}$, $\SI{94.47}{\mega\hertz}$ are remarkably similar to the DFT-calculated values ($\num{87}$, $\num{89}$, $\SI{92}{\mega\hertz}$, respectively) and have identical directions. A similar case is found for the nitrogen hyperfine, where the experimentally-measured values are within \SI{0.3}{\degree} of the DFT-calculated values. When taken in conjunction with the dopant and treatment history of \nfourtreated, these data are enough to conclusively assign the observed spectrum to the defect \SiVN [Fig~\ref{fig:orbitals}(a)].

The expected charge state of \SiVN can be calculated from the group theoretical descriptions of the SiV defects. Here, the neutral and negatively charged SiV defects possess two and one hole, respectively \cite{Gali2013}. In replacing one of the neighboring carbon atoms with nitrogen the number of holes present in the defect must decrease by one: we thus expect the positive, neutral and negative charge states to possess 2 ($S=0$ or 1), 1 ($S=1/2$), and zero ($S=0$) holes respectively, and we identify the new spectrum with \SiVNneutral{}. Charge transfer measurements on all samples are consistent with this description. The EPR spectrum is photochromic, with the concentration changing from approximately \num{400} to \SI{<5}{\ppb} between the UV and heat-treated charge states in \treatedtfh: the behavior of the defect is therefore qualitatively similar to the behavior of \SiVneutral{}. In the heated sample state, we observe no additional EPR spectra and deduce the dominant charge state is \SiVNminus{}, which is $S=0$ in its ground state and therefore EPR-inactive. 

\begin{table*}[bt]
	\centering
	\caption{Spin Hamiltonian parameters measured for \SiVNneutral{}. The three principal values ($p_{\numrange{1}{3}}$) and directions are given for each parameter. A positive tilt is given to mean away from \hkl[001] toward \hkl[110]: no tilt is required for the final principal value of each parameter, retaining the \hkl[1-10] mirror plane and reflecting the defect's \MonoclinicI{} symmetry.}
	\label{tab:spin_hamiltonian_parameters}
	\begin{tabular}{
		l
		l
		p{0.5cm}
		S[retain-explicit-plus,table-format=8.8]
		l
		S[retain-explicit-plus,table-format=+1.1]
		S[retain-explicit-plus,table-format=8.8]
		l
		S[retain-explicit-plus,table-format=+1.1]
		S[retain-explicit-plus,table-format=8.8]
		l
		S[retain-explicit-plus,table-format=+1.1]
		}
		\toprule
		
		Parameter 	& Unit 	&&
			\multicolumn{1}{c}{$p_1$} & \multicolumn{1}{c}{Dir.} &\multicolumn{1}{c}{Tilt (\si{\degree})}&
			\multicolumn{1}{c}{$p_2$} & \multicolumn{1}{c}{Dir.} & \multicolumn{1}{c}{Tilt (\si{\degree})}&
			\multicolumn{1}{c}{$p_3$} & \multicolumn{1}{c}{Dir.}\\
		\midrule
		$g$			&	1	& Exp. & 2.00472+-0.00005 & $\hkl[001]$ & +4.3 &
							   2.00549+-0.00005 & $\hkl[110]$ & +4.3 & 
							   2.00288+-0.00005 & $\hkl[1-10]$\\[0.8em]

		\multirow{2}{*}{$A$ ($^{14}\mathrm{N}$)} & \multirow{2}{*}{\si{\mega\hertz}} & Exp. & 
							   -3.800+-0.01 & $\hkl[111]$ & +2.7 &
							   -3.586+-0.01 & $\hkl[11-2]$ & +2.7 &
							   -3.281+-0.01 & $\hkl[1-10]$\\

			& 		& Theory&  -3.4			& $\hkl[111]$ & +6 &
							   -3.0			& $\hkl[11-2]$ & +2 &
							   -2.7 		& $\hkl[1-10]$\\[0.8em]

		$Q$ ($^{14}\mathrm{N}$)	& \si{\mega\hertz} & Exp. &
							   -2.078+-0.01 & $\hkl[111]$ & 0 &
							   +1.039+-0.01 & $\hkl[11-2]$ & 0 &
							   +1.039+-0.01 & $\hkl[1-10]$\\[0.8em]

		\multirow{2}{*}{$A$ ($^{29}\mathrm{Si}$)} & \multirow{2}{*}{\si{\mega\hertz}} & Exp. & 
								\pm98.24+-0.5 & $\hkl[22-1]$ & 0 &
								\pm98.13+-0.5 & $\hkl[114]$ & 0 &
								\pm94.47+-0.5 & $\hkl[1-10]$\\

			& 		& Theory&  +87 			 & $\hkl[22-1]$ & 8 &
							   +89			 & $\hkl[114]$  & 8 &
							   +92			 & $\hkl[1-10]$\\
		\bottomrule
	\end{tabular}
\end{table*}

DFT calculations of the stability of different charge states of \SiV and \SiVN are consistent with the observed charge state behavior: the neutral charge states of both defects are stable at approximately the same chemical potential, while \SiVNminus is the stable charge state over almost all other chemical potentials [Fig~\ref{fig:orbitals}(b)]. These calculations also predict that a double negatively charged \SiVN{} state can exist for high chemical potentials: this charge state would have one hole ($S=1/2$) and is able to form due to a disruption to the atomic configuration. Structurally, \SiVN{} can be compared to a \SiV system with a nitrogen donor and therefore \SiVNminus{} presents an electronically saturated system, as discussed above. DFT calculatons indicate that the addition of an extra electron to \SiVNminus{}, producing \SiVNtwominus{}, breaks a {$\mathrm{C}-\mathrm{N}$} bond with the nitrogen forming a \Ns structure bonded to the Si and two carbons. This geometric distortion to \SiVN results in the lowering of a band gap state which is now accesible for excitation. Examining the orbital characteristics depicted by spin density isosurfaces from DFT, we observe the \Nsneutral{}-like [Fig~\ref{fig:orbitals}(c)] configuration adopted by \SiVNtwominus{} [Fig~\ref{fig:orbitals}(d)], rather than retaining the configuration of the same band gap state in \SiVNneutral{} [Fig~\ref{fig:orbitals}(e)]. Calculations of the \SiVN charge stabilities [Table~\ref{tab:SiNV_binding_energies}] indicate that all three charge states are stable, and of these \SiVNminus is least likely to dissociate. 

\begin{table}[bh]
\centering
\caption{Binding energies ($E_\mathrm{bind}$) for each modeled defect through charge-conserving reactions. Displayed errors result from comparing values calculated using LDA and GGA functional.}
\begin{tabular}{lp{0.2cm}rclp{0.2cm}l}
\toprule
Defect && \multicolumn{3}{c}{Components} && \multicolumn{1}{r}{$E_\mathrm{bind}$ (eV)}\\
\midrule
\SiVNneutral{} 	&& \SiVminus{} &+& \NsPlus{} && \num{2.8+-0.01}  \\
\SiVNminus{} 	&& \SiVminus{} &+& \Nsneutral{} && \num{4.4+-0.1}  \\
\SiVNtwominus{} && \SiVtwominus{} &+& \Nsneutral{} && \num{1.8+-0.03}  \\
\bottomrule
\end{tabular}
\label{tab:SiNV_binding_energies}
\end{table}

We expect all charge states of the \SiVN to be difficult to identify in IR absorption measurements. The mass of the elements involved, combined with the vacancy, suggests that defect vibrations will be below the \SI{1332}{\per\centi\meter} lattice cutoff and therefore will contribute to the one-phonon absorption, rather than exhibiting sharp local vibrational modes. Unfortunately, the one-phonon IR absorption of samples~D--G contain other unidentified contributions thus no spectrum can be associated with any charge state of \SiVN at the present time.

\subsection{Defect production}
The addition of silicon (typically via silane) during CVD growth of diamond yields a grown-in (native) population of SiV centers \cite{Itoh2014,Rogers2014b}. In an analagous situation to nitrogen, where substitutional nitrogen concentrations are typically orders of magnitude higher than the grown-in NV concentrations \cite{Edmonds2012}, we presume the majority of the silicon is incorporated as substitutional silicon \cite{DHaenens-Johansson2011}, as discussed in \S\ref{subsec:annealing}. Therefore, there is a substantial source of silicon available within the sample itself from growth. 

We have not identified \SiVNneutral{} in any as-grown samples (putting an upper limit on the as-grown concentration of approximately \SI{0.5}{\ppb}). We first observe \SiVNneutral{} upon HPHT annealing at \SI{2000}{\celsius} (for one~hour, or \SI{100}{\hour} at \SI{1800}{\celsius}), and its concentration increases up to the maximum \SI{2400}{\celsius} temperature [Fig~\ref{fig:annealing}]. As is typical for vacancy-containing defects in diamond, we assume \SiVN{} production must occur via vacancy-assisted migration of impurities, as the energy required for direct diffusion of substitutional nitrogen (\SI{8}{\electronvolt} \cite{Goss2006}) and silicon is significantly higher than the vacancy-assisted mechanisms \cite{Mainwood1994}. Furthermore, the diffusion barrier for \NVminus (\SI{\approx5}{\electronvolt} \cite{Pinto2012}) is significantly lower than for \SiVneutral (\SI{\approx6.5}{\electronvolt} \cite{Gali2013,Goss2007}). At high temperatures where \NV is unstable, nitrogen may diffuse through the lattice by concerted exchange with a vacancy before the NV pair breaks up \cite{Pinto2012}. We therefore understand \SiVN production to occur via the diffusion of vacancies and subsequent capture by \Sis, producing \SiV; and the vacancy-assisted diffusion of nitrogen to \SiV centers producing \SiVN.

Recent reports of delayed luminescence at \SI{499}{\nano\meter} from synthetic, silicon-containing samples suggested that the emission originates at $\mathrm{Si}_x\mathrm{N}_y$ or $\mathrm{Si}_x\mathrm{N}_y\mathrm{V}$ complexes \cite{Wassell2018}. The \SI{499}{\nano\meter} luminescence is maximized on annealing at \SI{1700}{\celsius} and destroyed above \SI{2000}{\celsius} \cite{Wassell2018}. We do not observe this luminescence from any of the present samples at any annealing temperature. Additionally, the annealing behavior of \SiVN is incompatible with the reported annealing behavior of the \SI{499}{\nano\meter} defect and therefore we conclude that the luminescence does not originate at \SiVN. Furthermore, as \SiVN is the simplest variant of the $\mathrm{Si}_x\mathrm{N}_y\mathrm{V}$ defects, and defect aggregation in diamond typically develops from simple to more complex under higher annealing temperatures, it seems unlikely that the emission originates with any defect in this group. 

\section{Conclusion}
The present samples, while dominated in total concentration by nitrogen-related defects, enable additional insight into silicon-related defects and processes which must occur even in lower-concentration samples. The first observation of a doubly-charged defect in diamond leads the way for future studies of other doubly-charged donors or acceptors, provided the ($+$/$2+$) or ($-$/$2-$) levels are sufficiently deep.

The existence of \SiVtwominus{} puts limits on the production efficiency of \SiVminus{} qubits in nitrogen-doped material. Previous reports which interpreted the absence of \SiVminus in n-type material as the presence of \SiVneutral{} should now be re-interpreted in terms of charge transfer between \SiVtwominus{} and \SiVminus{}, rather than \SiVneutral{} and \SiVminus{} \cite{Dhomkar2018}. As UV light is expected to be required to directly ionize \SiVtwominus{} it is not clear that it will be possible to design a simple optical ionization protocol to drive $\SiVtwominus{}\rightarrow\SiVminus{}$ --- any pulse which ionizes \SiVtwominus{} is likely also to drive charge in other proximal defects, reducing overall charge stability of the ensemble. Devices which require \SiVminus{} as the dominant charge state should therefore be intrinsic or only moderately n-type to avoid interference from \SiVtwominus{}.

\begin{acknowledgments}
We are grateful to Ben Truscott (Element Six) for growing the samples, to John Freeth \& Hugh Leach (De Beers Group Technology) for preparing and polishing the samples, and to David Hardeman (Element Six) for annealing the samples. We thank Andrew Edmonds (Element Six) for careful reading of the manuscript. We acknowledge use of Spectroscopy Research Technology Platform facilities at the University of Warwick. B.\ L.\ G.\ acknowledges funding from the Royal Academy of Engineering.
\end{acknowledgments}


\begin{thebibliography}{79}%
\makeatletter
\providecommand \@ifxundefined [1]{%
 \@ifx{#1\undefined}
}%
\providecommand \@ifnum [1]{%
 \ifnum #1\expandafter \@firstoftwo
 \else \expandafter \@secondoftwo
 \fi
}%
\providecommand \@ifx [1]{%
 \ifx #1\expandafter \@firstoftwo
 \else \expandafter \@secondoftwo
 \fi
}%
\providecommand \natexlab [1]{#1}%
\providecommand \enquote  [1]{``#1''}%
\providecommand \bibnamefont  [1]{#1}%
\providecommand \bibfnamefont [1]{#1}%
\providecommand \citenamefont [1]{#1}%
\providecommand \href@noop [0]{\@secondoftwo}%
\providecommand \href [0]{\begingroup \@sanitize@url \@href}%
\providecommand \@href[1]{\@@startlink{#1}\@@href}%
\providecommand \@@href[1]{\endgroup#1\@@endlink}%
\providecommand \@sanitize@url [0]{\catcode `\\12\catcode `\$12\catcode
  `\&12\catcode `\#12\catcode `\^12\catcode `\_12\catcode `\%12\relax}%
\providecommand \@@startlink[1]{}%
\providecommand \@@endlink[0]{}%
\providecommand \url  [0]{\begingroup\@sanitize@url \@url }%
\providecommand \@url [1]{\endgroup\@href {#1}{\urlprefix }}%
\providecommand \urlprefix  [0]{URL }%
\providecommand \Eprint [0]{\href }%
\providecommand \doibase [0]{http://dx.doi.org/}%
\providecommand \selectlanguage [0]{\@gobble}%
\providecommand \bibinfo  [0]{\@secondoftwo}%
\providecommand \bibfield  [0]{\@secondoftwo}%
\providecommand \translation [1]{[#1]}%
\providecommand \BibitemOpen [0]{}%
\providecommand \bibitemStop [0]{}%
\providecommand \bibitemNoStop [0]{.\EOS\space}%
\providecommand \EOS [0]{\spacefactor3000\relax}%
\providecommand \BibitemShut  [1]{\csname bibitem#1\endcsname}%
\let\auto@bib@innerbib\@empty
\bibitem [{\citenamefont {Atat{\"{u}}re}\ \emph {et~al.}(2018)\citenamefont
  {Atat{\"{u}}re}, \citenamefont {Englund}, \citenamefont {Vamivakas},
  \citenamefont {Lee},\ and\ \citenamefont {Wrachtrup}}]{Atature2018}%
  \BibitemOpen
  \bibfield  {author} {\bibinfo {author} {\bibfnamefont {M.}~\bibnamefont
  {Atat{\"{u}}re}}, \bibinfo {author} {\bibfnamefont {D.}~\bibnamefont
  {Englund}}, \bibinfo {author} {\bibfnamefont {N.}~\bibnamefont {Vamivakas}},
  \bibinfo {author} {\bibfnamefont {S.-Y.}\ \bibnamefont {Lee}}, \ and\
  \bibinfo {author} {\bibfnamefont {J.}~\bibnamefont {Wrachtrup}},\ }\href
  {\doibase 10.1038/s41578-018-0008-9} {\bibfield  {journal} {\bibinfo
  {journal} {Nat. Rev. Mater.}\ }\textbf {\bibinfo {volume} {3}},\ \bibinfo
  {pages} {38} (\bibinfo {year} {2018})}\BibitemShut {NoStop}%
\bibitem [{\citenamefont {Aharonovich}\ and\ \citenamefont
  {Neu}(2014)}]{Aharonovich2014}%
  \BibitemOpen
  \bibfield  {author} {\bibinfo {author} {\bibfnamefont {I.}~\bibnamefont
  {Aharonovich}}\ and\ \bibinfo {author} {\bibfnamefont {E.}~\bibnamefont
  {Neu}},\ }\href {\doibase 10.1002/adom.201400189} {\bibfield  {journal}
  {\bibinfo  {journal} {Adv. Opt. Mater.}\ }\textbf {\bibinfo {volume} {2}},\
  \bibinfo {pages} {911} (\bibinfo {year} {2014})}\BibitemShut {NoStop}%
\bibitem [{\citenamefont {Schr{\"{o}}der}\ \emph {et~al.}(2016)\citenamefont
  {Schr{\"{o}}der}, \citenamefont {Mouradian}, \citenamefont {Zheng},
  \citenamefont {Trusheim}, \citenamefont {Walsh}, \citenamefont {Chen},
  \citenamefont {Li}, \citenamefont {Bayn},\ and\ \citenamefont
  {Englund}}]{Schroder2016}%
  \BibitemOpen
  \bibfield  {author} {\bibinfo {author} {\bibfnamefont {T.}~\bibnamefont
  {Schr{\"{o}}der}}, \bibinfo {author} {\bibfnamefont {S.~L.}\ \bibnamefont
  {Mouradian}}, \bibinfo {author} {\bibfnamefont {J.}~\bibnamefont {Zheng}},
  \bibinfo {author} {\bibfnamefont {M.~E.}\ \bibnamefont {Trusheim}}, \bibinfo
  {author} {\bibfnamefont {M.}~\bibnamefont {Walsh}}, \bibinfo {author}
  {\bibfnamefont {E.~H.}\ \bibnamefont {Chen}}, \bibinfo {author}
  {\bibfnamefont {L.}~\bibnamefont {Li}}, \bibinfo {author} {\bibfnamefont
  {I.}~\bibnamefont {Bayn}}, \ and\ \bibinfo {author} {\bibfnamefont
  {D.}~\bibnamefont {Englund}},\ }\href {\doibase 10.1364/JOSAB.33.000B65}
  {\bibfield  {journal} {\bibinfo  {journal} {J. Opt. Soc. Am. B}\ }\textbf
  {\bibinfo {volume} {33}},\ \bibinfo {pages} {B65} (\bibinfo {year}
  {2016})}\BibitemShut {NoStop}%
\bibitem [{\citenamefont {Awschalom}\ \emph {et~al.}(2018)\citenamefont
  {Awschalom}, \citenamefont {Hanson}, \citenamefont {Wrachtrup},\ and\
  \citenamefont {Zhou}}]{Awschalom2018}%
  \BibitemOpen
  \bibfield  {author} {\bibinfo {author} {\bibfnamefont {D.~D.}\ \bibnamefont
  {Awschalom}}, \bibinfo {author} {\bibfnamefont {R.}~\bibnamefont {Hanson}},
  \bibinfo {author} {\bibfnamefont {J.}~\bibnamefont {Wrachtrup}}, \ and\
  \bibinfo {author} {\bibfnamefont {B.~B.}\ \bibnamefont {Zhou}},\ }\href
  {\doibase 10.1038/s41566-018-0232-2} {\bibfield  {journal} {\bibinfo
  {journal} {Nat. Photonics}\ }\textbf {\bibinfo {volume} {12}},\ \bibinfo
  {pages} {516} (\bibinfo {year} {2018})}\BibitemShut {NoStop}%
\bibitem [{\citenamefont {Doherty}\ \emph {et~al.}(2013)\citenamefont
  {Doherty}, \citenamefont {Manson}, \citenamefont {Delaney}, \citenamefont
  {Jelezko}, \citenamefont {Wrachtrup},\ and\ \citenamefont
  {Hollenberg}}]{Doherty2013}%
  \BibitemOpen
  \bibfield  {author} {\bibinfo {author} {\bibfnamefont {M.~W.}\ \bibnamefont
  {Doherty}}, \bibinfo {author} {\bibfnamefont {N.~B.}\ \bibnamefont {Manson}},
  \bibinfo {author} {\bibfnamefont {P.}~\bibnamefont {Delaney}}, \bibinfo
  {author} {\bibfnamefont {F.}~\bibnamefont {Jelezko}}, \bibinfo {author}
  {\bibfnamefont {J.}~\bibnamefont {Wrachtrup}}, \ and\ \bibinfo {author}
  {\bibfnamefont {L.~C.~L.}\ \bibnamefont {Hollenberg}},\ }\href {\doibase
  10.1016/j.physrep.2013.02.001} {\bibfield  {journal} {\bibinfo  {journal}
  {Phys. Rep.}\ }\textbf {\bibinfo {volume} {528}},\ \bibinfo {pages} {1}
  (\bibinfo {year} {2013})}\BibitemShut {NoStop}%
\bibitem [{\citenamefont {Evans}\ \emph {et~al.}(2018)\citenamefont {Evans},
  \citenamefont {Bhaskar}, \citenamefont {Sukachev}, \citenamefont {Nguyen},
  \citenamefont {Sipahigil}, \citenamefont {Burek}, \citenamefont {Machielse},
  \citenamefont {Zhang}, \citenamefont {Zibrov}, \citenamefont {Bielejec},
  \citenamefont {Park}, \citenamefont {Lon{\v{c}}ar},\ and\ \citenamefont
  {Lukin}}]{Evans2018}%
  \BibitemOpen
  \bibfield  {author} {\bibinfo {author} {\bibfnamefont {R.~E.}\ \bibnamefont
  {Evans}}, \bibinfo {author} {\bibfnamefont {M.~K.}\ \bibnamefont {Bhaskar}},
  \bibinfo {author} {\bibfnamefont {D.~D.}\ \bibnamefont {Sukachev}}, \bibinfo
  {author} {\bibfnamefont {C.~T.}\ \bibnamefont {Nguyen}}, \bibinfo {author}
  {\bibfnamefont {A.}~\bibnamefont {Sipahigil}}, \bibinfo {author}
  {\bibfnamefont {M.~J.}\ \bibnamefont {Burek}}, \bibinfo {author}
  {\bibfnamefont {B.}~\bibnamefont {Machielse}}, \bibinfo {author}
  {\bibfnamefont {G.~H.}\ \bibnamefont {Zhang}}, \bibinfo {author}
  {\bibfnamefont {A.~S.}\ \bibnamefont {Zibrov}}, \bibinfo {author}
  {\bibfnamefont {E.}~\bibnamefont {Bielejec}}, \bibinfo {author}
  {\bibfnamefont {H.}~\bibnamefont {Park}}, \bibinfo {author} {\bibfnamefont
  {M.}~\bibnamefont {Lon{\v{c}}ar}}, \ and\ \bibinfo {author} {\bibfnamefont
  {M.~D.}\ \bibnamefont {Lukin}},\ }\href {\doibase 10.1126/science.aau4691}
  {\bibfield  {journal} {\bibinfo  {journal} {Science}\ }\textbf {\bibinfo
  {volume} {362}},\ \bibinfo {pages} {662} (\bibinfo {year}
  {2018})}\BibitemShut {NoStop}%
\bibitem [{\citenamefont {Rose}\ \emph {et~al.}(2018)\citenamefont {Rose},
  \citenamefont {Huang}, \citenamefont {Zhang}, \citenamefont {Stevenson},
  \citenamefont {Tyryshkin}, \citenamefont {Sangtawesin}, \citenamefont
  {Srinivasan}, \citenamefont {Loudin}, \citenamefont {Markham}, \citenamefont
  {Edmonds}, \citenamefont {Twitchen}, \citenamefont {Lyon},\ and\
  \citenamefont {de~Leon}}]{Rose2018}%
  \BibitemOpen
  \bibfield  {author} {\bibinfo {author} {\bibfnamefont {B.~C.}\ \bibnamefont
  {Rose}}, \bibinfo {author} {\bibfnamefont {D.}~\bibnamefont {Huang}},
  \bibinfo {author} {\bibfnamefont {Z.-H.}\ \bibnamefont {Zhang}}, \bibinfo
  {author} {\bibfnamefont {P.}~\bibnamefont {Stevenson}}, \bibinfo {author}
  {\bibfnamefont {A.~M.}\ \bibnamefont {Tyryshkin}}, \bibinfo {author}
  {\bibfnamefont {S.}~\bibnamefont {Sangtawesin}}, \bibinfo {author}
  {\bibfnamefont {S.}~\bibnamefont {Srinivasan}}, \bibinfo {author}
  {\bibfnamefont {L.}~\bibnamefont {Loudin}}, \bibinfo {author} {\bibfnamefont
  {M.~L.}\ \bibnamefont {Markham}}, \bibinfo {author} {\bibfnamefont {A.~M.}\
  \bibnamefont {Edmonds}}, \bibinfo {author} {\bibfnamefont {D.~J.}\
  \bibnamefont {Twitchen}}, \bibinfo {author} {\bibfnamefont {S.~A.}\
  \bibnamefont {Lyon}}, \ and\ \bibinfo {author} {\bibfnamefont {N.~P.}\
  \bibnamefont {de~Leon}},\ }\href {\doibase 10.1126/science.aao0290}
  {\bibfield  {journal} {\bibinfo  {journal} {Science}\ }\textbf {\bibinfo
  {volume} {361}},\ \bibinfo {pages} {60} (\bibinfo {year} {2018})}\BibitemShut
  {NoStop}%
\bibitem [{\citenamefont {Green}\ \emph
  {et~al.}(2017{\natexlab{a}})\citenamefont {Green}, \citenamefont {Mottishaw},
  \citenamefont {Breeze}, \citenamefont {Edmonds}, \citenamefont
  {D'Haenens-Johansson}, \citenamefont {Doherty}, \citenamefont {Williams},
  \citenamefont {Twitchen},\ and\ \citenamefont {Newton}}]{Green2017c}%
  \BibitemOpen
  \bibfield  {author} {\bibinfo {author} {\bibfnamefont {B.~L.}\ \bibnamefont
  {Green}}, \bibinfo {author} {\bibfnamefont {S.}~\bibnamefont {Mottishaw}},
  \bibinfo {author} {\bibfnamefont {B.~G.}\ \bibnamefont {Breeze}}, \bibinfo
  {author} {\bibfnamefont {A.~M.}\ \bibnamefont {Edmonds}}, \bibinfo {author}
  {\bibfnamefont {U.~F.~S.}\ \bibnamefont {D'Haenens-Johansson}}, \bibinfo
  {author} {\bibfnamefont {M.~W.}\ \bibnamefont {Doherty}}, \bibinfo {author}
  {\bibfnamefont {S.~D.}\ \bibnamefont {Williams}}, \bibinfo {author}
  {\bibfnamefont {D.~J.}\ \bibnamefont {Twitchen}}, \ and\ \bibinfo {author}
  {\bibfnamefont {M.~E.}\ \bibnamefont {Newton}},\ }\href {\doibase
  10.1103/PhysRevLett.119.096402} {\bibfield  {journal} {\bibinfo  {journal}
  {Phys. Rev. Lett.}\ }\textbf {\bibinfo {volume} {119}},\ \bibinfo {pages}
  {096402} (\bibinfo {year} {2017}{\natexlab{a}})}\BibitemShut {NoStop}%
\bibitem [{\citenamefont {Iwasaki}\ \emph {et~al.}(2017)\citenamefont
  {Iwasaki}, \citenamefont {Miyamoto}, \citenamefont {Taniguchi}, \citenamefont
  {Siyushev}, \citenamefont {Metsch}, \citenamefont {Jelezko},\ and\
  \citenamefont {Hatano}}]{Iwasaki2017}%
  \BibitemOpen
  \bibfield  {author} {\bibinfo {author} {\bibfnamefont {T.}~\bibnamefont
  {Iwasaki}}, \bibinfo {author} {\bibfnamefont {Y.}~\bibnamefont {Miyamoto}},
  \bibinfo {author} {\bibfnamefont {T.}~\bibnamefont {Taniguchi}}, \bibinfo
  {author} {\bibfnamefont {P.}~\bibnamefont {Siyushev}}, \bibinfo {author}
  {\bibfnamefont {M.~H.}\ \bibnamefont {Metsch}}, \bibinfo {author}
  {\bibfnamefont {F.}~\bibnamefont {Jelezko}}, \ and\ \bibinfo {author}
  {\bibfnamefont {M.}~\bibnamefont {Hatano}},\ }\href {\doibase
  10.1103/PhysRevLett.119.253601} {\bibfield  {journal} {\bibinfo  {journal}
  {Phys. Rev. Lett.}\ }\textbf {\bibinfo {volume} {119}},\ \bibinfo {pages}
  {253601} (\bibinfo {year} {2017})}\BibitemShut {NoStop}%
\bibitem [{\citenamefont {Trusheim}\ \emph {et~al.}(2019)\citenamefont
  {Trusheim}, \citenamefont {Wan}, \citenamefont {Chen}, \citenamefont
  {Ciccarino}, \citenamefont {Flick}, \citenamefont {Sundararaman},
  \citenamefont {Malladi}, \citenamefont {Bersin}, \citenamefont {Walsh},
  \citenamefont {Lienhard}, \citenamefont {Bakhru}, \citenamefont {Narang},\
  and\ \citenamefont {Englund}}]{Trusheim2019}%
  \BibitemOpen
  \bibfield  {author} {\bibinfo {author} {\bibfnamefont {M.~E.}\ \bibnamefont
  {Trusheim}}, \bibinfo {author} {\bibfnamefont {N.~H.}\ \bibnamefont {Wan}},
  \bibinfo {author} {\bibfnamefont {K.~C.}\ \bibnamefont {Chen}}, \bibinfo
  {author} {\bibfnamefont {C.~J.}\ \bibnamefont {Ciccarino}}, \bibinfo {author}
  {\bibfnamefont {J.}~\bibnamefont {Flick}}, \bibinfo {author} {\bibfnamefont
  {R.}~\bibnamefont {Sundararaman}}, \bibinfo {author} {\bibfnamefont
  {G.}~\bibnamefont {Malladi}}, \bibinfo {author} {\bibfnamefont
  {E.}~\bibnamefont {Bersin}}, \bibinfo {author} {\bibfnamefont
  {M.}~\bibnamefont {Walsh}}, \bibinfo {author} {\bibfnamefont
  {B.}~\bibnamefont {Lienhard}}, \bibinfo {author} {\bibfnamefont
  {H.}~\bibnamefont {Bakhru}}, \bibinfo {author} {\bibfnamefont
  {P.}~\bibnamefont {Narang}}, \ and\ \bibinfo {author} {\bibfnamefont
  {D.}~\bibnamefont {Englund}},\ }\href {\doibase 10.1103/PhysRevB.99.075430}
  {\bibfield  {journal} {\bibinfo  {journal} {Phys. Rev. B}\ }\textbf {\bibinfo
  {volume} {99}},\ \bibinfo {pages} {075430} (\bibinfo {year}
  {2019})}\BibitemShut {NoStop}%
\bibitem [{\citenamefont {Zaitsev}(2001)}]{Zaitsev2001}%
  \BibitemOpen
  \bibfield  {author} {\bibinfo {author} {\bibfnamefont {A.~M.}\ \bibnamefont
  {Zaitsev}},\ }\href@noop {} {\emph {\bibinfo {title} {{Optical Properties of
  Diamond}}}}\ (\bibinfo  {publisher} {Springer},\ \bibinfo {year}
  {2001})\BibitemShut {NoStop}%
\bibitem [{\citenamefont {Dischler}(2012)}]{Dischler2012}%
  \BibitemOpen
  \bibfield  {author} {\bibinfo {author} {\bibfnamefont {B.}~\bibnamefont
  {Dischler}},\ }\href@noop {} {\emph {\bibinfo {title} {{Handbook of Spectral
  Lines in Diamond}}}}\ (\bibinfo  {publisher} {Springer-Verlag},\ \bibinfo
  {address} {Berlin},\ \bibinfo {year} {2012})\BibitemShut {NoStop}%
\bibitem [{\citenamefont {Goss}\ \emph {et~al.}(1996)\citenamefont {Goss},
  \citenamefont {Jones}, \citenamefont {Breuer}, \citenamefont {Briddon},\ and\
  \citenamefont {{\"O}berg}}]{Goss1996}%
  \BibitemOpen
  \bibfield  {author} {\bibinfo {author} {\bibfnamefont {J.~P.}\ \bibnamefont
  {Goss}}, \bibinfo {author} {\bibfnamefont {R.}~\bibnamefont {Jones}},
  \bibinfo {author} {\bibfnamefont {S.~J.}\ \bibnamefont {Breuer}}, \bibinfo
  {author} {\bibfnamefont {P.~R.}\ \bibnamefont {Briddon}}, \ and\ \bibinfo
  {author} {\bibfnamefont {S.}~\bibnamefont {{\"O}berg}},\ }\href
  {http://link.aps.org/doi/10.1103/PhysRevLett.77.3041} {\bibfield  {journal}
  {\bibinfo  {journal} {Phys. Rev. Lett.}\ }\textbf {\bibinfo {volume} {77}},\
  \bibinfo {pages} {3041} (\bibinfo {year} {1996})}\BibitemShut {NoStop}%
\bibitem [{\citenamefont {Rogers}\ \emph
  {et~al.}(2014{\natexlab{a}})\citenamefont {Rogers}, \citenamefont {Jahnke},
  \citenamefont {Doherty}, \citenamefont {Dietrich}, \citenamefont
  {McGuinness}, \citenamefont {M{\"{u}}ller}, \citenamefont {Teraji},
  \citenamefont {Sumiya}, \citenamefont {Isoya}, \citenamefont {Manson},\ and\
  \citenamefont {Jelezko}}]{Rogers2014a}%
  \BibitemOpen
  \bibfield  {author} {\bibinfo {author} {\bibfnamefont {L.~J.}\ \bibnamefont
  {Rogers}}, \bibinfo {author} {\bibfnamefont {K.~D.}\ \bibnamefont {Jahnke}},
  \bibinfo {author} {\bibfnamefont {M.~W.}\ \bibnamefont {Doherty}}, \bibinfo
  {author} {\bibfnamefont {A.}~\bibnamefont {Dietrich}}, \bibinfo {author}
  {\bibfnamefont {L.~P.}\ \bibnamefont {McGuinness}}, \bibinfo {author}
  {\bibfnamefont {C.}~\bibnamefont {M{\"{u}}ller}}, \bibinfo {author}
  {\bibfnamefont {T.}~\bibnamefont {Teraji}}, \bibinfo {author} {\bibfnamefont
  {H.}~\bibnamefont {Sumiya}}, \bibinfo {author} {\bibfnamefont
  {J.}~\bibnamefont {Isoya}}, \bibinfo {author} {\bibfnamefont {N.~B.}\
  \bibnamefont {Manson}}, \ and\ \bibinfo {author} {\bibfnamefont
  {F.}~\bibnamefont {Jelezko}},\ }\href {\doibase 10.1103/PhysRevB.89.235101}
  {\bibfield  {journal} {\bibinfo  {journal} {Phys. Rev. B}\ }\textbf {\bibinfo
  {volume} {89}},\ \bibinfo {pages} {235101} (\bibinfo {year}
  {2014}{\natexlab{a}})}\BibitemShut {NoStop}%
\bibitem [{\citenamefont {Hepp}\ \emph {et~al.}(2014)\citenamefont {Hepp},
  \citenamefont {M{\"{u}}ller}, \citenamefont {Waselowski}, \citenamefont
  {Becker}, \citenamefont {Pingault}, \citenamefont {Sternschulte},
  \citenamefont {Steinm{\"{u}}ller-Nethl}, \citenamefont {Gali}, \citenamefont
  {Maze}, \citenamefont {Atat{\"{u}}re},\ and\ \citenamefont
  {Becher}}]{Hepp2014a}%
  \BibitemOpen
  \bibfield  {author} {\bibinfo {author} {\bibfnamefont {C.}~\bibnamefont
  {Hepp}}, \bibinfo {author} {\bibfnamefont {T.}~\bibnamefont {M{\"{u}}ller}},
  \bibinfo {author} {\bibfnamefont {V.}~\bibnamefont {Waselowski}}, \bibinfo
  {author} {\bibfnamefont {J.~N.}\ \bibnamefont {Becker}}, \bibinfo {author}
  {\bibfnamefont {B.}~\bibnamefont {Pingault}}, \bibinfo {author}
  {\bibfnamefont {H.}~\bibnamefont {Sternschulte}}, \bibinfo {author}
  {\bibfnamefont {D.}~\bibnamefont {Steinm{\"{u}}ller-Nethl}}, \bibinfo
  {author} {\bibfnamefont {A.}~\bibnamefont {Gali}}, \bibinfo {author}
  {\bibfnamefont {J.~R.}\ \bibnamefont {Maze}}, \bibinfo {author}
  {\bibfnamefont {M.}~\bibnamefont {Atat{\"{u}}re}}, \ and\ \bibinfo {author}
  {\bibfnamefont {C.}~\bibnamefont {Becher}},\ }\href {\doibase
  10.1103/PhysRevLett.112.036405} {\bibfield  {journal} {\bibinfo  {journal}
  {Phys. Rev. Lett.}\ }\textbf {\bibinfo {volume} {112}},\ \bibinfo {pages}
  {036405} (\bibinfo {year} {2014})}\BibitemShut {NoStop}%
\bibitem [{\citenamefont {D'Haenens-Johansson}\ \emph
  {et~al.}(2011)\citenamefont {D'Haenens-Johansson}, \citenamefont {Edmonds},
  \citenamefont {Green}, \citenamefont {Newton}, \citenamefont {Davies},
  \citenamefont {Martineau}, \citenamefont {Khan},\ and\ \citenamefont
  {Twitchen}}]{DHaenens-Johansson2011}%
  \BibitemOpen
  \bibfield  {author} {\bibinfo {author} {\bibfnamefont {U.~F.}\ \bibnamefont
  {D'Haenens-Johansson}}, \bibinfo {author} {\bibfnamefont {A.~M.}\
  \bibnamefont {Edmonds}}, \bibinfo {author} {\bibfnamefont {B.~L.}\
  \bibnamefont {Green}}, \bibinfo {author} {\bibfnamefont {M.~E.}\ \bibnamefont
  {Newton}}, \bibinfo {author} {\bibfnamefont {G.}~\bibnamefont {Davies}},
  \bibinfo {author} {\bibfnamefont {P.~M.}\ \bibnamefont {Martineau}}, \bibinfo
  {author} {\bibfnamefont {R.~U.}\ \bibnamefont {Khan}}, \ and\ \bibinfo
  {author} {\bibfnamefont {D.~J.}\ \bibnamefont {Twitchen}},\ }\href {\doibase
  10.1103/PhysRevB.84.245208} {\bibfield  {journal} {\bibinfo  {journal} {Phys.
  Rev. B - Condens. Matter Mater. Phys.}\ }\textbf {\bibinfo {volume} {84}},\
  \bibinfo {pages} {245208} (\bibinfo {year} {2011})}\BibitemShut {NoStop}%
\bibitem [{\citenamefont {Green}\ \emph {et~al.}(2019)\citenamefont {Green},
  \citenamefont {Doherty}, \citenamefont {Nako}, \citenamefont {Manson},
  \citenamefont {D'Haenens-Johansson}, \citenamefont {Williams}, \citenamefont
  {Twitchen},\ and\ \citenamefont {Newton}}]{Green2019}%
  \BibitemOpen
  \bibfield  {author} {\bibinfo {author} {\bibfnamefont {B.~L.}\ \bibnamefont
  {Green}}, \bibinfo {author} {\bibfnamefont {M.~W.}\ \bibnamefont {Doherty}},
  \bibinfo {author} {\bibfnamefont {E.}~\bibnamefont {Nako}}, \bibinfo {author}
  {\bibfnamefont {N.~B.}\ \bibnamefont {Manson}}, \bibinfo {author}
  {\bibfnamefont {U.~F.~S.}\ \bibnamefont {D'Haenens-Johansson}}, \bibinfo
  {author} {\bibfnamefont {S.~D.}\ \bibnamefont {Williams}}, \bibinfo {author}
  {\bibfnamefont {D.~J.}\ \bibnamefont {Twitchen}}, \ and\ \bibinfo {author}
  {\bibfnamefont {M.~E.}\ \bibnamefont {Newton}},\ }\href {\doibase
  10.1103/PhysRevB.99.161112} {\bibfield  {journal} {\bibinfo  {journal} {Phys.
  Rev. B}\ }\textbf {\bibinfo {volume} {99}},\ \bibinfo {pages} {161112(R)}
  (\bibinfo {year} {2019})}\BibitemShut {NoStop}%
\bibitem [{\citenamefont {Iakoubovskii}\ \emph {et~al.}(2003)\citenamefont
  {Iakoubovskii}, \citenamefont {Stesmans}, \citenamefont {Suzuki},
  \citenamefont {Kuwabara},\ and\ \citenamefont {Sawabe}}]{Iakoubovskii2003a}%
  \BibitemOpen
  \bibfield  {author} {\bibinfo {author} {\bibfnamefont {K.}~\bibnamefont
  {Iakoubovskii}}, \bibinfo {author} {\bibfnamefont {A.}~\bibnamefont
  {Stesmans}}, \bibinfo {author} {\bibfnamefont {K.}~\bibnamefont {Suzuki}},
  \bibinfo {author} {\bibfnamefont {J.}~\bibnamefont {Kuwabara}}, \ and\
  \bibinfo {author} {\bibfnamefont {A.}~\bibnamefont {Sawabe}},\ }\href
  {\doibase 10.1016/S0925-9635(02)00380-1} {\bibfield  {journal} {\bibinfo
  {journal} {Diam. Relat. Mater.}\ }\textbf {\bibinfo {volume} {12}},\ \bibinfo
  {pages} {511} (\bibinfo {year} {2003})}\BibitemShut {NoStop}%
\bibitem [{\citenamefont {Edmonds}\ \emph {et~al.}(2008)\citenamefont
  {Edmonds}, \citenamefont {Newton}, \citenamefont {Martineau}, \citenamefont
  {Twitchen},\ and\ \citenamefont {Williams}}]{Edmonds2008a}%
  \BibitemOpen
  \bibfield  {author} {\bibinfo {author} {\bibfnamefont {A.~M.}\ \bibnamefont
  {Edmonds}}, \bibinfo {author} {\bibfnamefont {M.~E.}\ \bibnamefont {Newton}},
  \bibinfo {author} {\bibfnamefont {P.~M.}\ \bibnamefont {Martineau}}, \bibinfo
  {author} {\bibfnamefont {D.~J.}\ \bibnamefont {Twitchen}}, \ and\ \bibinfo
  {author} {\bibfnamefont {S.~D.}\ \bibnamefont {Williams}},\ }\href {\doibase
  10.1103/PhysRevB.77.245205} {\bibfield  {journal} {\bibinfo  {journal} {Phys.
  Rev. B}\ }\textbf {\bibinfo {volume} {77}},\ \bibinfo {pages} {245205}
  (\bibinfo {year} {2008})}\BibitemShut {NoStop}%
\bibitem [{\citenamefont {D'Haenens-Johansson}\ \emph
  {et~al.}(2010)\citenamefont {D'Haenens-Johansson}, \citenamefont {Edmonds},
  \citenamefont {Newton}, \citenamefont {Goss}, \citenamefont {Briddon},
  \citenamefont {Baker}, \citenamefont {Martineau}, \citenamefont {Khan},
  \citenamefont {Twitchen},\ and\ \citenamefont
  {Williams}}]{DHaenens-Johansson2010}%
  \BibitemOpen
  \bibfield  {author} {\bibinfo {author} {\bibfnamefont {U.~F.~S.}\
  \bibnamefont {D'Haenens-Johansson}}, \bibinfo {author} {\bibfnamefont
  {A.~M.}\ \bibnamefont {Edmonds}}, \bibinfo {author} {\bibfnamefont {M.~E.}\
  \bibnamefont {Newton}}, \bibinfo {author} {\bibfnamefont {J.~P.}\
  \bibnamefont {Goss}}, \bibinfo {author} {\bibfnamefont {P.~R.}\ \bibnamefont
  {Briddon}}, \bibinfo {author} {\bibfnamefont {J.~M.}\ \bibnamefont {Baker}},
  \bibinfo {author} {\bibfnamefont {P.~M.}\ \bibnamefont {Martineau}}, \bibinfo
  {author} {\bibfnamefont {R.~U.~A.}\ \bibnamefont {Khan}}, \bibinfo {author}
  {\bibfnamefont {D.~J.}\ \bibnamefont {Twitchen}}, \ and\ \bibinfo {author}
  {\bibfnamefont {S.~D.}\ \bibnamefont {Williams}},\ }\href {\doibase
  10.1103/PhysRevB.82.155205} {\bibfield  {journal} {\bibinfo  {journal} {Phys.
  Rev. B}\ }\textbf {\bibinfo {volume} {82}},\ \bibinfo {pages} {155205}
  (\bibinfo {year} {2010})}\BibitemShut {NoStop}%
\bibitem [{\citenamefont {Nadolinny}\ \emph {et~al.}(2016)\citenamefont
  {Nadolinny}, \citenamefont {Komarovskikh}, \citenamefont {Palyanov},
  \citenamefont {Kupriyanov}, \citenamefont {Borzdov}, \citenamefont
  {Rakhmanova}, \citenamefont {Yuryeva},\ and\ \citenamefont
  {Veber}}]{Nadolinny2016}%
  \BibitemOpen
  \bibfield  {author} {\bibinfo {author} {\bibfnamefont {V.}~\bibnamefont
  {Nadolinny}}, \bibinfo {author} {\bibfnamefont {A.}~\bibnamefont
  {Komarovskikh}}, \bibinfo {author} {\bibfnamefont {Y.}~\bibnamefont
  {Palyanov}}, \bibinfo {author} {\bibfnamefont {I.}~\bibnamefont
  {Kupriyanov}}, \bibinfo {author} {\bibfnamefont {Y.}~\bibnamefont {Borzdov}},
  \bibinfo {author} {\bibfnamefont {M.}~\bibnamefont {Rakhmanova}}, \bibinfo
  {author} {\bibfnamefont {O.}~\bibnamefont {Yuryeva}}, \ and\ \bibinfo
  {author} {\bibfnamefont {S.}~\bibnamefont {Veber}},\ }\href {\doibase
  10.1002/pssa.201600211} {\bibfield  {journal} {\bibinfo  {journal} {Phys.
  Status Solidi Appl. Mater. Sci.}\ }\textbf {\bibinfo {volume} {213}},\
  \bibinfo {pages} {2623} (\bibinfo {year} {2016})}\BibitemShut {NoStop}%
\bibitem [{\citenamefont {Goss}\ \emph {et~al.}(2007)\citenamefont {Goss},
  \citenamefont {Briddon},\ and\ \citenamefont {Shaw}}]{Goss2007}%
  \BibitemOpen
  \bibfield  {author} {\bibinfo {author} {\bibfnamefont {J.~P.}\ \bibnamefont
  {Goss}}, \bibinfo {author} {\bibfnamefont {P.~R.}\ \bibnamefont {Briddon}}, \
  and\ \bibinfo {author} {\bibfnamefont {M.~J.}\ \bibnamefont {Shaw}},\ }\href
  {\doibase 10.1103/PhysRevB.76.075204} {\bibfield  {journal} {\bibinfo
  {journal} {Phys. Rev. B}\ }\textbf {\bibinfo {volume} {76}},\ \bibinfo
  {pages} {1} (\bibinfo {year} {2007})}\BibitemShut {NoStop}%
\bibitem [{\citenamefont {Thiering}\ and\ \citenamefont
  {Gali}(2015)}]{Thiering2015}%
  \BibitemOpen
  \bibfield  {author} {\bibinfo {author} {\bibfnamefont {G.}~\bibnamefont
  {Thiering}}\ and\ \bibinfo {author} {\bibfnamefont {A.}~\bibnamefont
  {Gali}},\ }\href {\doibase 10.1103/PhysRevB.92.165203} {\bibfield  {journal}
  {\bibinfo  {journal} {Phys. Rev. B - Condens. Matter Mater. Phys.}\ }\textbf
  {\bibinfo {volume} {92}},\ \bibinfo {pages} {165203} (\bibinfo {year}
  {2015})}\BibitemShut {NoStop}%
\bibitem [{\citenamefont {Khan}\ \emph {et~al.}(2009)\citenamefont {Khan},
  \citenamefont {Martineau}, \citenamefont {Cann}, \citenamefont {Newton},\
  and\ \citenamefont {Twitchen}}]{Khan2009}%
  \BibitemOpen
  \bibfield  {author} {\bibinfo {author} {\bibfnamefont {R.~U.~A.}\
  \bibnamefont {Khan}}, \bibinfo {author} {\bibfnamefont {P.~M.}\ \bibnamefont
  {Martineau}}, \bibinfo {author} {\bibfnamefont {B.~L.}\ \bibnamefont {Cann}},
  \bibinfo {author} {\bibfnamefont {M.~E.}\ \bibnamefont {Newton}}, \ and\
  \bibinfo {author} {\bibfnamefont {D.~J.}\ \bibnamefont {Twitchen}},\ }\href
  {\doibase 10.1088/0953-8984/21/36/364214} {\bibfield  {journal} {\bibinfo
  {journal} {J. Phys. Condens. Matter}\ }\textbf {\bibinfo {volume} {21}},\
  \bibinfo {pages} {364214} (\bibinfo {year} {2009})}\BibitemShut {NoStop}%
\bibitem [{\citenamefont {Khan}\ \emph {et~al.}(2013)\citenamefont {Khan},
  \citenamefont {Cann}, \citenamefont {Martineau}, \citenamefont {Samartseva},
  \citenamefont {Freeth}, \citenamefont {Sibley}, \citenamefont {Hartland},
  \citenamefont {Newton}, \citenamefont {Dhillon},\ and\ \citenamefont
  {Twitchen}}]{Khan2013}%
  \BibitemOpen
  \bibfield  {author} {\bibinfo {author} {\bibfnamefont {R.~U.~A.}\
  \bibnamefont {Khan}}, \bibinfo {author} {\bibfnamefont {B.~L.}\ \bibnamefont
  {Cann}}, \bibinfo {author} {\bibfnamefont {P.~M.}\ \bibnamefont {Martineau}},
  \bibinfo {author} {\bibfnamefont {J.}~\bibnamefont {Samartseva}}, \bibinfo
  {author} {\bibfnamefont {J.~J.~P.}\ \bibnamefont {Freeth}}, \bibinfo {author}
  {\bibfnamefont {S.~J.}\ \bibnamefont {Sibley}}, \bibinfo {author}
  {\bibfnamefont {C.~B.}\ \bibnamefont {Hartland}}, \bibinfo {author}
  {\bibfnamefont {M.~E.}\ \bibnamefont {Newton}}, \bibinfo {author}
  {\bibfnamefont {H.~K.}\ \bibnamefont {Dhillon}}, \ and\ \bibinfo {author}
  {\bibfnamefont {D.~J.}\ \bibnamefont {Twitchen}},\ }\href {\doibase
  10.1088/0953-8984/25/27/275801} {\bibfield  {journal} {\bibinfo  {journal}
  {J. Phys. Condens. Matter}\ }\textbf {\bibinfo {volume} {25}},\ \bibinfo
  {pages} {275801} (\bibinfo {year} {2013})}\BibitemShut {NoStop}%
\bibitem [{\citenamefont {M{\"{a}}ki}\ \emph {et~al.}(2011)\citenamefont
  {M{\"{a}}ki}, \citenamefont {Tuomisto}, \citenamefont {Varpula},
  \citenamefont {Fisher}, \citenamefont {Khan},\ and\ \citenamefont
  {Martineau}}]{Maki2011}%
  \BibitemOpen
  \bibfield  {author} {\bibinfo {author} {\bibfnamefont {J.-M.}\ \bibnamefont
  {M{\"{a}}ki}}, \bibinfo {author} {\bibfnamefont {F.}~\bibnamefont
  {Tuomisto}}, \bibinfo {author} {\bibfnamefont {A.}~\bibnamefont {Varpula}},
  \bibinfo {author} {\bibfnamefont {D.}~\bibnamefont {Fisher}}, \bibinfo
  {author} {\bibfnamefont {R.~U.~A.}\ \bibnamefont {Khan}}, \ and\ \bibinfo
  {author} {\bibfnamefont {P.~M.}\ \bibnamefont {Martineau}},\ }\href {\doibase
  10.1103/PhysRevLett.107.217403} {\bibfield  {journal} {\bibinfo  {journal}
  {Phys. Rev. Lett.}\ }\textbf {\bibinfo {volume} {107}},\ \bibinfo {pages}
  {217403} (\bibinfo {year} {2011})}\BibitemShut {NoStop}%
\bibitem [{\citenamefont {Dale}(2015)}]{Dale2015}%
  \BibitemOpen
  \bibfield  {author} {\bibinfo {author} {\bibfnamefont {M.~W.}\ \bibnamefont
  {Dale}},\ }\emph {\bibinfo {title} {{Colour centres on demand in diamond}}},\
  \href@noop {} {Ph.D. thesis},\ \bibinfo  {school} {University of Warwick}
  (\bibinfo {year} {2015})\BibitemShut {NoStop}%
\bibitem [{\citenamefont {Green}\ \emph
  {et~al.}(2017{\natexlab{b}})\citenamefont {Green}, \citenamefont {Breeze},\
  and\ \citenamefont {Newton}}]{Green2017}%
  \BibitemOpen
  \bibfield  {author} {\bibinfo {author} {\bibfnamefont {B.~L.}\ \bibnamefont
  {Green}}, \bibinfo {author} {\bibfnamefont {B.~G.}\ \bibnamefont {Breeze}}, \
  and\ \bibinfo {author} {\bibfnamefont {M.~E.}\ \bibnamefont {Newton}},\
  }\href {\doibase 10.1088/1361-648X/aa6c89} {\bibfield  {journal} {\bibinfo
  {journal} {J. Phys. Condens. Matter}\ }\textbf {\bibinfo {volume} {29}},\
  \bibinfo {pages} {225701} (\bibinfo {year} {2017}{\natexlab{b}})}\BibitemShut
  {NoStop}%
\bibitem [{\citenamefont {Davies}(1999)}]{Davies1999}%
  \BibitemOpen
  \bibfield  {author} {\bibinfo {author} {\bibfnamefont {G.}~\bibnamefont
  {Davies}},\ }\href
  {http://www.sciencedirect.com/science/article/B6TVH-3YYVCK0-5Y/2/483f0c8b713a875e2ea3c55668a0996c}
  {\bibfield  {journal} {\bibinfo  {journal} {Phys. B Condens. Matter}\
  }\textbf {\bibinfo {volume} {273-274}},\ \bibinfo {pages} {15} (\bibinfo
  {year} {1999})}\BibitemShut {NoStop}%
\bibitem [{\citenamefont {Rayson}\ and\ \citenamefont
  {Briddon}(2009)}]{rayson2009highly}%
  \BibitemOpen
  \bibfield  {author} {\bibinfo {author} {\bibfnamefont {M.~J.}\ \bibnamefont
  {Rayson}}\ and\ \bibinfo {author} {\bibfnamefont {P.~R.}\ \bibnamefont
  {Briddon}},\ }\href@noop {} {\bibfield  {journal} {\bibinfo  {journal}
  {Physical Review B}\ }\textbf {\bibinfo {volume} {80}},\ \bibinfo {pages}
  {205104} (\bibinfo {year} {2009})}\BibitemShut {NoStop}%
\bibitem [{\citenamefont {Perdew}\ \emph {et~al.}(1996)\citenamefont {Perdew},
  \citenamefont {Burke},\ and\ \citenamefont
  {Ernzerhof}}]{perdew1996generalized}%
  \BibitemOpen
  \bibfield  {author} {\bibinfo {author} {\bibfnamefont {J.~P.}\ \bibnamefont
  {Perdew}}, \bibinfo {author} {\bibfnamefont {K.}~\bibnamefont {Burke}}, \
  and\ \bibinfo {author} {\bibfnamefont {M.}~\bibnamefont {Ernzerhof}},\
  }\href@noop {} {\bibfield  {journal} {\bibinfo  {journal} {Physical review
  letters}\ }\textbf {\bibinfo {volume} {77}},\ \bibinfo {pages} {3865}
  (\bibinfo {year} {1996})}\BibitemShut {NoStop}%
\bibitem [{\citenamefont {Hartwigsen}\ \emph {et~al.}(1998)\citenamefont
  {Hartwigsen}, \citenamefont {G{\oe}decker},\ and\ \citenamefont
  {Hutter}}]{hartwigsen1998relativistic}%
  \BibitemOpen
  \bibfield  {author} {\bibinfo {author} {\bibfnamefont {C.}~\bibnamefont
  {Hartwigsen}}, \bibinfo {author} {\bibfnamefont {S.}~\bibnamefont
  {G{\oe}decker}}, \ and\ \bibinfo {author} {\bibfnamefont {J.}~\bibnamefont
  {Hutter}},\ }\href@noop {} {\bibfield  {journal} {\bibinfo  {journal}
  {Physical Review B}\ }\textbf {\bibinfo {volume} {58}},\ \bibinfo {pages}
  {3641} (\bibinfo {year} {1998})}\BibitemShut {NoStop}%
\bibitem [{\citenamefont {Monkhorst}\ and\ \citenamefont
  {Pack}(1976)}]{monkhorst1976special}%
  \BibitemOpen
  \bibfield  {author} {\bibinfo {author} {\bibfnamefont {H.~J.}\ \bibnamefont
  {Monkhorst}}\ and\ \bibinfo {author} {\bibfnamefont {J.~D.}\ \bibnamefont
  {Pack}},\ }\href@noop {} {\bibfield  {journal} {\bibinfo  {journal} {Physical
  review B}\ }\textbf {\bibinfo {volume} {13}},\ \bibinfo {pages} {5188}
  (\bibinfo {year} {1976})}\BibitemShut {NoStop}%
\bibitem [{\citenamefont {Riley}(1944)}]{riley1944lattice}%
  \BibitemOpen
  \bibfield  {author} {\bibinfo {author} {\bibfnamefont {D.~P.}\ \bibnamefont
  {Riley}},\ }\href@noop {} {\bibfield  {journal} {\bibinfo  {journal}
  {Nature}\ }\textbf {\bibinfo {volume} {153}},\ \bibinfo {pages} {587}
  (\bibinfo {year} {1944})}\BibitemShut {NoStop}%
\bibitem [{\citenamefont {Goss}\ \emph {et~al.}(2004)\citenamefont {Goss},
  \citenamefont {Briddon}, \citenamefont {Jones},\ and\ \citenamefont
  {Sque}}]{goss2004}%
  \BibitemOpen
  \bibfield  {author} {\bibinfo {author} {\bibfnamefont {J.~P.}\ \bibnamefont
  {Goss}}, \bibinfo {author} {\bibfnamefont {P.~R.}\ \bibnamefont {Briddon}},
  \bibinfo {author} {\bibfnamefont {R.}~\bibnamefont {Jones}}, \ and\ \bibinfo
  {author} {\bibfnamefont {S.}~\bibnamefont {Sque}},\ }\href@noop {} {\bibfield
   {journal} {\bibinfo  {journal} {Diamond and related materials}\ }\textbf
  {\bibinfo {volume} {13}},\ \bibinfo {pages} {684} (\bibinfo {year}
  {2004})}\BibitemShut {NoStop}%
\bibitem [{\citenamefont {Zhang}\ and\ \citenamefont {Northrup}(1991)}]{chemp}%
  \BibitemOpen
  \bibfield  {author} {\bibinfo {author} {\bibfnamefont {S.~B.}\ \bibnamefont
  {Zhang}}\ and\ \bibinfo {author} {\bibfnamefont {J.~E.}\ \bibnamefont
  {Northrup}},\ }\href@noop {} {\bibfield  {journal} {\bibinfo  {journal}
  {Phys. Rev. Lett.}\ }\textbf {\bibinfo {volume} {67}},\ \bibinfo {pages}
  {2339} (\bibinfo {year} {1991})}\BibitemShut {NoStop}%
\bibitem [{\citenamefont {Shim}\ \emph {et~al.}(2005)\citenamefont {Shim},
  \citenamefont {Lee}, \citenamefont {Lee},\ and\ \citenamefont
  {Nieminen}}]{shim2005density}%
  \BibitemOpen
  \bibfield  {author} {\bibinfo {author} {\bibfnamefont {J.}~\bibnamefont
  {Shim}}, \bibinfo {author} {\bibfnamefont {E.-K.}\ \bibnamefont {Lee}},
  \bibinfo {author} {\bibfnamefont {Y.}~\bibnamefont {Lee}}, \ and\ \bibinfo
  {author} {\bibfnamefont {R.~M.}\ \bibnamefont {Nieminen}},\ }\href@noop {}
  {\bibfield  {journal} {\bibinfo  {journal} {Physical Review B}\ }\textbf
  {\bibinfo {volume} {71}},\ \bibinfo {pages} {035206} (\bibinfo {year}
  {2005})}\BibitemShut {NoStop}%
\bibitem [{\citenamefont {Goss}\ and\ \citenamefont
  {Briddon}(2008)}]{goss2008dissociation}%
  \BibitemOpen
  \bibfield  {author} {\bibinfo {author} {\bibfnamefont {J.~P.}\ \bibnamefont
  {Goss}}\ and\ \bibinfo {author} {\bibfnamefont {P.~R.}\ \bibnamefont
  {Briddon}},\ }\href@noop {} {\bibfield  {journal} {\bibinfo  {journal}
  {Physical Review B}\ }\textbf {\bibinfo {volume} {77}},\ \bibinfo {pages}
  {035211} (\bibinfo {year} {2008})}\BibitemShut {NoStop}%
\bibitem [{\citenamefont {Shaw}\ \emph {et~al.}(2005)\citenamefont {Shaw},
  \citenamefont {Briddon}, \citenamefont {Goss}, \citenamefont {Rayson},
  \citenamefont {Kerridge}, \citenamefont {Harker},\ and\ \citenamefont
  {Stoneham}}]{shaw2005importance}%
  \BibitemOpen
  \bibfield  {author} {\bibinfo {author} {\bibfnamefont {M.~J.}\ \bibnamefont
  {Shaw}}, \bibinfo {author} {\bibfnamefont {P.~R.}\ \bibnamefont {Briddon}},
  \bibinfo {author} {\bibfnamefont {J.~P.}\ \bibnamefont {Goss}}, \bibinfo
  {author} {\bibfnamefont {M.~J.}\ \bibnamefont {Rayson}}, \bibinfo {author}
  {\bibfnamefont {A.}~\bibnamefont {Kerridge}}, \bibinfo {author}
  {\bibfnamefont {A.~H.}\ \bibnamefont {Harker}}, \ and\ \bibinfo {author}
  {\bibfnamefont {A.~M.}\ \bibnamefont {Stoneham}},\ }\href@noop {} {\bibfield
  {journal} {\bibinfo  {journal} {Physical review letters}\ }\textbf {\bibinfo
  {volume} {95}},\ \bibinfo {pages} {105502} (\bibinfo {year}
  {2005})}\BibitemShut {NoStop}%
\bibitem [{\citenamefont {Woods}\ \emph {et~al.}(1990)\citenamefont {Woods},
  \citenamefont {{Van Wyk}},\ and\ \citenamefont {Collins}}]{Woods1990}%
  \BibitemOpen
  \bibfield  {author} {\bibinfo {author} {\bibfnamefont {G.~S.}\ \bibnamefont
  {Woods}}, \bibinfo {author} {\bibfnamefont {J.~A.}\ \bibnamefont {{Van
  Wyk}}}, \ and\ \bibinfo {author} {\bibfnamefont {A.~T.}\ \bibnamefont
  {Collins}},\ }\href {\doibase 10.1080/13642819008215257} {\bibfield
  {journal} {\bibinfo  {journal} {Philos. Mag. Part B}\ }\textbf {\bibinfo
  {volume} {62}},\ \bibinfo {pages} {589} (\bibinfo {year} {1990})}\BibitemShut
  {NoStop}%
\bibitem [{\citenamefont {Lawson}\ \emph {et~al.}(1998)\citenamefont {Lawson},
  \citenamefont {Fisher}, \citenamefont {Hunt},\ and\ \citenamefont
  {Newton}}]{Lawson1998}%
  \BibitemOpen
  \bibfield  {author} {\bibinfo {author} {\bibfnamefont {S.~C.}\ \bibnamefont
  {Lawson}}, \bibinfo {author} {\bibfnamefont {D.}~\bibnamefont {Fisher}},
  \bibinfo {author} {\bibfnamefont {D.~C.}\ \bibnamefont {Hunt}}, \ and\
  \bibinfo {author} {\bibfnamefont {M.~E.}\ \bibnamefont {Newton}},\ }\href
  {\doibase 10.1088/0953-8984/10/27/016} {\bibfield  {journal} {\bibinfo
  {journal} {J. Phys. Condens. Matter}\ }\textbf {\bibinfo {volume} {10}},\
  \bibinfo {pages} {6171} (\bibinfo {year} {1998})}\BibitemShut {NoStop}%
\bibitem [{\citenamefont {Liggins}(2010)}]{Liggins2010b}%
  \BibitemOpen
  \bibfield  {author} {\bibinfo {author} {\bibfnamefont {S.}~\bibnamefont
  {Liggins}},\ }\emph {\bibinfo {title} {{Identification of point defects in
  treated single crystal diamond}}},\ \href
  {http://webcat.warwick.ac.uk/record=b2491628{~}S15} {Ph.D. thesis},\ \bibinfo
   {school} {University of Warwick} (\bibinfo {year} {2010})\BibitemShut
  {NoStop}%
\bibitem [{\citenamefont {Glover}\ \emph {et~al.}(2003)\citenamefont {Glover},
  \citenamefont {Newton}, \citenamefont {Martineau}, \citenamefont {Twitchen},\
  and\ \citenamefont {Baker}}]{Glover2003a}%
  \BibitemOpen
  \bibfield  {author} {\bibinfo {author} {\bibfnamefont {C.}~\bibnamefont
  {Glover}}, \bibinfo {author} {\bibfnamefont {M.}~\bibnamefont {Newton}},
  \bibinfo {author} {\bibfnamefont {P.}~\bibnamefont {Martineau}}, \bibinfo
  {author} {\bibfnamefont {D.}~\bibnamefont {Twitchen}}, \ and\ \bibinfo
  {author} {\bibfnamefont {J.}~\bibnamefont {Baker}},\ }\href {\doibase
  10.1103/PhysRevLett.90.185507} {\bibfield  {journal} {\bibinfo  {journal}
  {Phys. Rev. Lett.}\ }\textbf {\bibinfo {volume} {90}},\ \bibinfo {pages}
  {185507} (\bibinfo {year} {2003})}\BibitemShut {NoStop}%
\bibitem [{\citenamefont {Hartland}(2014)}]{Hartland2014}%
  \BibitemOpen
  \bibfield  {author} {\bibinfo {author} {\bibfnamefont {C.~B.}\ \bibnamefont
  {Hartland}},\ }\emph {\bibinfo {title} {{A Study of Point Defects in CVD
  Diamond Using Electron Paramagnetic Resonance and Optical Spectroscopy}}},\
  \href@noop {} {Ph.D. thesis},\ \bibinfo  {school} {University of Warwick}
  (\bibinfo {year} {2014})\BibitemShut {NoStop}%
\bibitem [{\citenamefont {Evans}\ and\ \citenamefont {Qi}(1982)}]{Evans1982}%
  \BibitemOpen
  \bibfield  {author} {\bibinfo {author} {\bibfnamefont {T.}~\bibnamefont
  {Evans}}\ and\ \bibinfo {author} {\bibfnamefont {Z.}~\bibnamefont {Qi}},\
  }\href
  {http://links.jstor.org/sici?sici=0080-4630(19820508)381:1780{\%}3C159:TKOTAO{\%}3E2.0.CO;2-V}
  {\bibfield  {journal} {\bibinfo  {journal} {Proc. R. Soc. London Ser. A}\
  }\textbf {\bibinfo {volume} {381}},\ \bibinfo {pages} {159} (\bibinfo {year}
  {1982})}\BibitemShut {NoStop}%
\bibitem [{\citenamefont {Wang}\ \emph {et~al.}(2007)\citenamefont {Wang},
  \citenamefont {Hall}, \citenamefont {Moe}, \citenamefont {Tower},\ and\
  \citenamefont {Moses}}]{Wang2007}%
  \BibitemOpen
  \bibfield  {author} {\bibinfo {author} {\bibfnamefont {W.}~\bibnamefont
  {Wang}}, \bibinfo {author} {\bibfnamefont {M.~S.}\ \bibnamefont {Hall}},
  \bibinfo {author} {\bibfnamefont {K.~S.}\ \bibnamefont {Moe}}, \bibinfo
  {author} {\bibfnamefont {J.}~\bibnamefont {Tower}}, \ and\ \bibinfo {author}
  {\bibfnamefont {T.~M.}\ \bibnamefont {Moses}},\ }\href {\doibase
  10.5741/GEMS.43.4.294} {\bibfield  {journal} {\bibinfo  {journal} {Gems
  Gemol.}\ }\textbf {\bibinfo {volume} {43}},\ \bibinfo {pages} {294} (\bibinfo
  {year} {2007})}\BibitemShut {NoStop}%
\bibitem [{\citenamefont {Khan}\ \emph {et~al.}(2010)\citenamefont {Khan},
  \citenamefont {Martineau}, \citenamefont {Cann}, \citenamefont {Newton},
  \citenamefont {Dhillon},\ and\ \citenamefont {Twitchen}}]{Khan2010}%
  \BibitemOpen
  \bibfield  {author} {\bibinfo {author} {\bibfnamefont {R.~U.~A.}\
  \bibnamefont {Khan}}, \bibinfo {author} {\bibfnamefont {P.~M.}\ \bibnamefont
  {Martineau}}, \bibinfo {author} {\bibfnamefont {B.~L.}\ \bibnamefont {Cann}},
  \bibinfo {author} {\bibfnamefont {M.~E.}\ \bibnamefont {Newton}}, \bibinfo
  {author} {\bibfnamefont {H.~K.}\ \bibnamefont {Dhillon}}, \ and\ \bibinfo
  {author} {\bibfnamefont {D.~J.}\ \bibnamefont {Twitchen}},\ }\href@noop {}
  {\bibfield  {journal} {\bibinfo  {journal} {Gems Gemol.}\ }\textbf {\bibinfo
  {volume} {46}},\ \bibinfo {pages} {18} (\bibinfo {year} {2010})}\BibitemShut
  {NoStop}%
\bibitem [{\citenamefont {Palyanov}\ \emph {et~al.}(2017)\citenamefont
  {Palyanov}, \citenamefont {Kupriyanov}, \citenamefont {Borzdov},
  \citenamefont {Nechaev},\ and\ \citenamefont {Bataleva}}]{Palyanov2017}%
  \BibitemOpen
  \bibfield  {author} {\bibinfo {author} {\bibfnamefont {Y.}~\bibnamefont
  {Palyanov}}, \bibinfo {author} {\bibfnamefont {I.}~\bibnamefont
  {Kupriyanov}}, \bibinfo {author} {\bibfnamefont {Y.}~\bibnamefont {Borzdov}},
  \bibinfo {author} {\bibfnamefont {D.}~\bibnamefont {Nechaev}}, \ and\
  \bibinfo {author} {\bibfnamefont {Y.}~\bibnamefont {Bataleva}},\ }\href
  {\doibase 10.3390/cryst7050119} {\bibfield  {journal} {\bibinfo  {journal}
  {Crystals}\ }\textbf {\bibinfo {volume} {7}},\ \bibinfo {pages} {119}
  (\bibinfo {year} {2017})}\BibitemShut {NoStop}%
\bibitem [{\citenamefont {Palyanov}\ \emph {et~al.}(2015)\citenamefont
  {Palyanov}, \citenamefont {Kupriyanov}, \citenamefont {Borzdov},\ and\
  \citenamefont {Bataleva}}]{Palyanov2015b}%
  \BibitemOpen
  \bibfield  {author} {\bibinfo {author} {\bibfnamefont {Y.~N.}\ \bibnamefont
  {Palyanov}}, \bibinfo {author} {\bibfnamefont {I.~N.}\ \bibnamefont
  {Kupriyanov}}, \bibinfo {author} {\bibfnamefont {Y.~M.}\ \bibnamefont
  {Borzdov}}, \ and\ \bibinfo {author} {\bibfnamefont {Y.~V.}\ \bibnamefont
  {Bataleva}},\ }\href {\doibase 10.1039/c5ce01265a} {\bibfield  {journal}
  {\bibinfo  {journal} {CrystEngComm}\ }\textbf {\bibinfo {volume} {17}},\
  \bibinfo {pages} {7323} (\bibinfo {year} {2015})}\BibitemShut {NoStop}%
\bibitem [{\citenamefont {Martineau}\ \emph {et~al.}(2004)\citenamefont
  {Martineau}, \citenamefont {Lawson}, \citenamefont {Taylor}, \citenamefont
  {Quinn}, \citenamefont {Evans},\ and\ \citenamefont
  {Crowder}}]{Martineau2004}%
  \BibitemOpen
  \bibfield  {author} {\bibinfo {author} {\bibfnamefont {P.}~\bibnamefont
  {Martineau}}, \bibinfo {author} {\bibfnamefont {S.~C.}\ \bibnamefont
  {Lawson}}, \bibinfo {author} {\bibfnamefont {A.~J.}\ \bibnamefont {Taylor}},
  \bibinfo {author} {\bibfnamefont {S.~J.}\ \bibnamefont {Quinn}}, \bibinfo
  {author} {\bibfnamefont {D.~J.~F.}\ \bibnamefont {Evans}}, \ and\ \bibinfo
  {author} {\bibfnamefont {M.}~\bibnamefont {Crowder}},\ }\href@noop {}
  {\bibfield  {journal} {\bibinfo  {journal} {Gems Gemol.}\ }\textbf {\bibinfo
  {volume} {40}},\ \bibinfo {pages} {2} (\bibinfo {year} {2004})}\BibitemShut
  {NoStop}%
\bibitem [{\citenamefont {H{\"{a}}u{\ss}ler}\ \emph {et~al.}(2017)\citenamefont
  {H{\"{a}}u{\ss}ler}, \citenamefont {Thiering}, \citenamefont {Dietrich},
  \citenamefont {Waasem}, \citenamefont {Teraji}, \citenamefont {Isoya},
  \citenamefont {Iwasaki}, \citenamefont {Hatano}, \citenamefont {Jelezko},
  \citenamefont {Gali},\ and\ \citenamefont {Kubanek}}]{Haußler2017}%
  \BibitemOpen
  \bibfield  {author} {\bibinfo {author} {\bibfnamefont {S.}~\bibnamefont
  {H{\"{a}}u{\ss}ler}}, \bibinfo {author} {\bibfnamefont {G.}~\bibnamefont
  {Thiering}}, \bibinfo {author} {\bibfnamefont {A.}~\bibnamefont {Dietrich}},
  \bibinfo {author} {\bibfnamefont {N.}~\bibnamefont {Waasem}}, \bibinfo
  {author} {\bibfnamefont {T.}~\bibnamefont {Teraji}}, \bibinfo {author}
  {\bibfnamefont {J.}~\bibnamefont {Isoya}}, \bibinfo {author} {\bibfnamefont
  {T.}~\bibnamefont {Iwasaki}}, \bibinfo {author} {\bibfnamefont
  {M.}~\bibnamefont {Hatano}}, \bibinfo {author} {\bibfnamefont
  {F.}~\bibnamefont {Jelezko}}, \bibinfo {author} {\bibfnamefont
  {A.}~\bibnamefont {Gali}}, \ and\ \bibinfo {author} {\bibfnamefont
  {A.}~\bibnamefont {Kubanek}},\ }\href {\doibase 10.1088/1367-2630/aa73e5}
  {\bibfield  {journal} {\bibinfo  {journal} {New J. Phys.}\ }\textbf {\bibinfo
  {volume} {19}},\ \bibinfo {pages} {063036} (\bibinfo {year}
  {2017})}\BibitemShut {NoStop}%
\bibitem [{\citenamefont {Ekimov}\ \emph {et~al.}(2017)\citenamefont {Ekimov},
  \citenamefont {Krivobok}, \citenamefont {Lyapin}, \citenamefont {Sherin},
  \citenamefont {Gavva},\ and\ \citenamefont {Kondrin}}]{Ekimov2017a}%
  \BibitemOpen
  \bibfield  {author} {\bibinfo {author} {\bibfnamefont {E.~A.}\ \bibnamefont
  {Ekimov}}, \bibinfo {author} {\bibfnamefont {V.~S.}\ \bibnamefont
  {Krivobok}}, \bibinfo {author} {\bibfnamefont {S.~G.}\ \bibnamefont
  {Lyapin}}, \bibinfo {author} {\bibfnamefont {P.~S.}\ \bibnamefont {Sherin}},
  \bibinfo {author} {\bibfnamefont {V.~A.}\ \bibnamefont {Gavva}}, \ and\
  \bibinfo {author} {\bibfnamefont {M.~V.}\ \bibnamefont {Kondrin}},\ }\href
  {\doibase 10.1103/PhysRevB.95.094113} {\bibfield  {journal} {\bibinfo
  {journal} {Phys. Rev. B}\ }\textbf {\bibinfo {volume} {95}},\ \bibinfo
  {pages} {094113} (\bibinfo {year} {2017})}\BibitemShut {NoStop}%
\bibitem [{\citenamefont {Ekimov}\ \emph {et~al.}(2018)\citenamefont {Ekimov},
  \citenamefont {Sherin}, \citenamefont {Krivobok}, \citenamefont {Lyapin},
  \citenamefont {Gavva},\ and\ \citenamefont {Kondrin}}]{Ekimov2018}%
  \BibitemOpen
  \bibfield  {author} {\bibinfo {author} {\bibfnamefont {E.~A.}\ \bibnamefont
  {Ekimov}}, \bibinfo {author} {\bibfnamefont {P.~S.}\ \bibnamefont {Sherin}},
  \bibinfo {author} {\bibfnamefont {V.~S.}\ \bibnamefont {Krivobok}}, \bibinfo
  {author} {\bibfnamefont {S.~G.}\ \bibnamefont {Lyapin}}, \bibinfo {author}
  {\bibfnamefont {V.~A.}\ \bibnamefont {Gavva}}, \ and\ \bibinfo {author}
  {\bibfnamefont {M.~V.}\ \bibnamefont {Kondrin}},\ }\href {\doibase
  10.1103/PhysRevB.97.045206} {\bibfield  {journal} {\bibinfo  {journal} {Phys.
  Rev. B}\ }\textbf {\bibinfo {volume} {97}},\ \bibinfo {pages} {045206}
  (\bibinfo {year} {2018})}\BibitemShut {NoStop}%
\bibitem [{\citenamefont {Gali}\ and\ \citenamefont {Maze}(2013)}]{Gali2013}%
  \BibitemOpen
  \bibfield  {author} {\bibinfo {author} {\bibfnamefont {A.}~\bibnamefont
  {Gali}}\ and\ \bibinfo {author} {\bibfnamefont {J.~R.}\ \bibnamefont
  {Maze}},\ }\href {\doibase 10.1103/PhysRevB.88.235205} {\bibfield  {journal}
  {\bibinfo  {journal} {Phys. Rev. B}\ }\textbf {\bibinfo {volume} {88}},\
  \bibinfo {pages} {235205} (\bibinfo {year} {2013})}\BibitemShut {NoStop}%
\bibitem [{\citenamefont {Watkins}(2000)}]{Watkins2000}%
  \BibitemOpen
  \bibfield  {author} {\bibinfo {author} {\bibfnamefont {G.~D.}\ \bibnamefont
  {Watkins}},\ }\href {\doibase 10.1016/S1369-8001(00)00037-8} {\bibfield
  {journal} {\bibinfo  {journal} {Mater. Sci. Semicond. Process.}\ }\textbf
  {\bibinfo {volume} {3}},\ \bibinfo {pages} {227} (\bibinfo {year}
  {2000})}\BibitemShut {NoStop}%
\bibitem [{\citenamefont {Newman}(1982)}]{Newman1982}%
  \BibitemOpen
  \bibfield  {author} {\bibinfo {author} {\bibfnamefont {R.~C.}\ \bibnamefont
  {Newman}},\ }\href {\doibase 10.1088/0034-4885/45/10/003} {\bibfield
  {journal} {\bibinfo  {journal} {Reports Prog. Phys.}\ }\textbf {\bibinfo
  {volume} {45}},\ \bibinfo {pages} {1163} (\bibinfo {year}
  {1982})}\BibitemShut {NoStop}%
\bibitem [{\citenamefont {Weber}\ \emph {et~al.}(2013)\citenamefont {Weber},
  \citenamefont {Janotti},\ and\ \citenamefont {{Van De Walle}}}]{Weber2013}%
  \BibitemOpen
  \bibfield  {author} {\bibinfo {author} {\bibfnamefont {J.~R.}\ \bibnamefont
  {Weber}}, \bibinfo {author} {\bibfnamefont {A.}~\bibnamefont {Janotti}}, \
  and\ \bibinfo {author} {\bibfnamefont {C.~G.}\ \bibnamefont {{Van De
  Walle}}},\ }\href {\doibase 10.1103/PhysRevB.87.035203} {\bibfield  {journal}
  {\bibinfo  {journal} {Phys. Rev. B - Condens. Matter Mater. Phys.}\ }\textbf
  {\bibinfo {volume} {87}},\ \bibinfo {pages} {035203} (\bibinfo {year}
  {2013})}\BibitemShut {NoStop}%
\bibitem [{\citenamefont {Weber}\ \emph {et~al.}(2015)\citenamefont {Weber},
  \citenamefont {Janotti},\ and\ \citenamefont {{Van de Walle}}}]{Weber2015}%
  \BibitemOpen
  \bibfield  {author} {\bibinfo {author} {\bibfnamefont {J.~R.}\ \bibnamefont
  {Weber}}, \bibinfo {author} {\bibfnamefont {A.}~\bibnamefont {Janotti}}, \
  and\ \bibinfo {author} {\bibfnamefont {C.~G.}\ \bibnamefont {{Van de
  Walle}}},\ }in\ \href@noop {} {\emph {\bibinfo {booktitle} {Photonics
  Electron. with Ger.}}},\ \bibinfo {editor} {edited by\ \bibinfo {editor}
  {\bibfnamefont {K.}~\bibnamefont {Wada}}\ and\ \bibinfo {editor}
  {\bibfnamefont {L.~C.}\ \bibnamefont {Kimerling}}}\ (\bibinfo  {publisher}
  {Wiley-VCH Verlag},\ \bibinfo {year} {2015})\ \bibinfo {edition} {1st}\
  ed.\BibitemShut {Stop}%
\bibitem [{\citenamefont {Breuer}\ and\ \citenamefont
  {Briddon}(1995)}]{Breuer1995}%
  \BibitemOpen
  \bibfield  {author} {\bibinfo {author} {\bibfnamefont {S.}~\bibnamefont
  {Breuer}}\ and\ \bibinfo {author} {\bibfnamefont {P.}~\bibnamefont
  {Briddon}},\ }\href {\doibase 10.1103/PhysRevB.51.6984} {\bibfield  {journal}
  {\bibinfo  {journal} {Phys. Rev. B}\ }\textbf {\bibinfo {volume} {51}},\
  \bibinfo {pages} {6984} (\bibinfo {year} {1995})}\BibitemShut {NoStop}%
\bibitem [{\citenamefont {Mainwood}\ and\ \citenamefont
  {Stoneham}(1997)}]{Mainwood1997}%
  \BibitemOpen
  \bibfield  {author} {\bibinfo {author} {\bibfnamefont {A.}~\bibnamefont
  {Mainwood}}\ and\ \bibinfo {author} {\bibfnamefont {A.~M.}\ \bibnamefont
  {Stoneham}},\ }\href {\doibase 10.1088/0953-8984/9/11/013} {\bibfield
  {journal} {\bibinfo  {journal} {J. Phys. Condens. Matter}\ }\textbf {\bibinfo
  {volume} {9}},\ \bibinfo {pages} {2453} (\bibinfo {year} {1997})}\BibitemShut
  {NoStop}%
\bibitem [{\citenamefont {Watkins}\ and\ \citenamefont
  {Troxell}(1980)}]{Watkins1980}%
  \BibitemOpen
  \bibfield  {author} {\bibinfo {author} {\bibfnamefont {G.~D.}\ \bibnamefont
  {Watkins}}\ and\ \bibinfo {author} {\bibfnamefont {J.~R.}\ \bibnamefont
  {Troxell}},\ }\href {\doibase 10.1103/PhysRevLett.44.593} {\bibfield
  {journal} {\bibinfo  {journal} {Phys. Rev. Lett.}\ }\textbf {\bibinfo
  {volume} {44}},\ \bibinfo {pages} {593} (\bibinfo {year} {1980})}\BibitemShut
  {NoStop}%
\bibitem [{\citenamefont {Farrer}(1969)}]{Farrer1969a}%
  \BibitemOpen
  \bibfield  {author} {\bibinfo {author} {\bibfnamefont {R.~G.}\ \bibnamefont
  {Farrer}},\ }\href
  {http://www.sciencedirect.com/science/article/B6TVW-46X9MJF-1BT/2/6bda9932f4af5e431bf3f51cac6a482b}
  {\bibfield  {journal} {\bibinfo  {journal} {Solid State Commun.}\ }\textbf
  {\bibinfo {volume} {7}},\ \bibinfo {pages} {685} (\bibinfo {year}
  {1969})}\BibitemShut {NoStop}%
\bibitem [{\citenamefont {Jones}\ \emph {et~al.}(2009)\citenamefont {Jones},
  \citenamefont {Goss},\ and\ \citenamefont {Briddon}}]{Jones2009b}%
  \BibitemOpen
  \bibfield  {author} {\bibinfo {author} {\bibfnamefont {R.}~\bibnamefont
  {Jones}}, \bibinfo {author} {\bibfnamefont {J.~P.}\ \bibnamefont {Goss}}, \
  and\ \bibinfo {author} {\bibfnamefont {P.~R.}\ \bibnamefont {Briddon}},\
  }\href {\doibase 10.1103/PhysRevB.80.033205} {\bibfield  {journal} {\bibinfo
  {journal} {Phys. Rev. B}\ }\textbf {\bibinfo {volume} {80}},\ \bibinfo
  {pages} {033205} (\bibinfo {year} {2009})}\BibitemShut {NoStop}%
\bibitem [{\citenamefont {Kern}\ \emph {et~al.}(2017)\citenamefont {Kern},
  \citenamefont {Jeske}, \citenamefont {Lau}, \citenamefont {Greentree},
  \citenamefont {Jelezko},\ and\ \citenamefont {Twamley}}]{Kern2017}%
  \BibitemOpen
  \bibfield  {author} {\bibinfo {author} {\bibfnamefont {M.}~\bibnamefont
  {Kern}}, \bibinfo {author} {\bibfnamefont {J.}~\bibnamefont {Jeske}},
  \bibinfo {author} {\bibfnamefont {D.~W.}\ \bibnamefont {Lau}}, \bibinfo
  {author} {\bibfnamefont {A.~D.}\ \bibnamefont {Greentree}}, \bibinfo {author}
  {\bibfnamefont {F.}~\bibnamefont {Jelezko}}, \ and\ \bibinfo {author}
  {\bibfnamefont {J.}~\bibnamefont {Twamley}},\ }\href {\doibase
  10.1103/PhysRevB.95.235306} {\bibfield  {journal} {\bibinfo  {journal} {Phys.
  Rev. B}\ }\textbf {\bibinfo {volume} {95}},\ \bibinfo {pages} {1} (\bibinfo
  {year} {2017})}\BibitemShut {NoStop}%
\bibitem [{\citenamefont {Green}\ \emph {et~al.}(2015)\citenamefont {Green},
  \citenamefont {Dale}, \citenamefont {Newton},\ and\ \citenamefont
  {Fisher}}]{Green2015}%
  \BibitemOpen
  \bibfield  {author} {\bibinfo {author} {\bibfnamefont {B.~L.}\ \bibnamefont
  {Green}}, \bibinfo {author} {\bibfnamefont {M.~W.}\ \bibnamefont {Dale}},
  \bibinfo {author} {\bibfnamefont {M.~E.}\ \bibnamefont {Newton}}, \ and\
  \bibinfo {author} {\bibfnamefont {D.}~\bibnamefont {Fisher}},\ }\href
  {\doibase 10.1103/PhysRevB.92.165204} {\bibfield  {journal} {\bibinfo
  {journal} {Phys. Rev. B}\ }\textbf {\bibinfo {volume} {92}},\ \bibinfo
  {pages} {165204} (\bibinfo {year} {2015})}\BibitemShut {NoStop}%
\bibitem [{\citenamefont {Smith}\ \emph {et~al.}(1959)\citenamefont {Smith},
  \citenamefont {Sorokin}, \citenamefont {Gelles},\ and\ \citenamefont
  {Lasher}}]{Smith1959a}%
  \BibitemOpen
  \bibfield  {author} {\bibinfo {author} {\bibfnamefont {W.}~\bibnamefont
  {Smith}}, \bibinfo {author} {\bibfnamefont {P.}~\bibnamefont {Sorokin}},
  \bibinfo {author} {\bibfnamefont {I.}~\bibnamefont {Gelles}}, \ and\ \bibinfo
  {author} {\bibfnamefont {G.}~\bibnamefont {Lasher}},\ }\href {\doibase
  10.1103/PhysRev.115.1546} {\bibfield  {journal} {\bibinfo  {journal} {Phys.
  Rev.}\ }\textbf {\bibinfo {volume} {115}},\ \bibinfo {pages} {1546} (\bibinfo
  {year} {1959})}\BibitemShut {NoStop}%
\bibitem [{\citenamefont {Loubser}\ and\ \citenamefont {van
  Wyk}(1977)}]{Loubser1977}%
  \BibitemOpen
  \bibfield  {author} {\bibinfo {author} {\bibfnamefont {J.~H.~N.}\
  \bibnamefont {Loubser}}\ and\ \bibinfo {author} {\bibfnamefont {J.~A.}\
  \bibnamefont {van Wyk}},\ }\href@noop {} {\bibfield  {journal} {\bibinfo
  {journal} {Diam. Res.}\ }\textbf {\bibinfo {volume} {11}},\ \bibinfo {pages}
  {11} (\bibinfo {year} {1977})}\BibitemShut {NoStop}%
\bibitem [{\citenamefont {Loubser}\ and\ \citenamefont {van
  Wyk}(1978)}]{Loubser1978}%
  \BibitemOpen
  \bibfield  {author} {\bibinfo {author} {\bibfnamefont {J.~H.~N.}\
  \bibnamefont {Loubser}}\ and\ \bibinfo {author} {\bibfnamefont {J.~A.}\
  \bibnamefont {van Wyk}},\ }\href {\doibase 10.1088/0034-4885/41/8/002}
  {\bibfield  {journal} {\bibinfo  {journal} {Reports Prog. Phys.}\ }\textbf
  {\bibinfo {volume} {41}},\ \bibinfo {pages} {1201} (\bibinfo {year}
  {1978})}\BibitemShut {NoStop}%
\bibitem [{\citenamefont {Felton}\ \emph {et~al.}(2009)\citenamefont {Felton},
  \citenamefont {Cann}, \citenamefont {Edmonds}, \citenamefont {Liggins},
  \citenamefont {Cruddace}, \citenamefont {Newton}, \citenamefont {Fisher},\
  and\ \citenamefont {Baker}}]{Felton2009}%
  \BibitemOpen
  \bibfield  {author} {\bibinfo {author} {\bibfnamefont {S.}~\bibnamefont
  {Felton}}, \bibinfo {author} {\bibfnamefont {B.~L.}\ \bibnamefont {Cann}},
  \bibinfo {author} {\bibfnamefont {A.~M.}\ \bibnamefont {Edmonds}}, \bibinfo
  {author} {\bibfnamefont {S.}~\bibnamefont {Liggins}}, \bibinfo {author}
  {\bibfnamefont {R.~J.}\ \bibnamefont {Cruddace}}, \bibinfo {author}
  {\bibfnamefont {M.~E.}\ \bibnamefont {Newton}}, \bibinfo {author}
  {\bibfnamefont {D.}~\bibnamefont {Fisher}}, \ and\ \bibinfo {author}
  {\bibfnamefont {J.~M.}\ \bibnamefont {Baker}},\ }\href {\doibase
  10.1088/0953-8984/21/36/364212} {\bibfield  {journal} {\bibinfo  {journal}
  {J. Phys. Condens. Matter}\ }\textbf {\bibinfo {volume} {21}},\ \bibinfo
  {pages} {364212} (\bibinfo {year} {2009})}\BibitemShut {NoStop}%
\bibitem [{\citenamefont {Nadolinny}\ \emph {et~al.}(1999)\citenamefont
  {Nadolinny}, \citenamefont {Yelisseyev}, \citenamefont {Baker}, \citenamefont
  {Twitchen}, \citenamefont {Newton}, \citenamefont {Hofstaetter},\ and\
  \citenamefont {Feigelson}}]{Nadolinny1999a}%
  \BibitemOpen
  \bibfield  {author} {\bibinfo {author} {\bibfnamefont {V.}~\bibnamefont
  {Nadolinny}}, \bibinfo {author} {\bibfnamefont {A.}~\bibnamefont
  {Yelisseyev}}, \bibinfo {author} {\bibfnamefont {J.~M.}\ \bibnamefont
  {Baker}}, \bibinfo {author} {\bibfnamefont {D.}~\bibnamefont {Twitchen}},
  \bibinfo {author} {\bibfnamefont {M.~E.}\ \bibnamefont {Newton}}, \bibinfo
  {author} {\bibfnamefont {A.}~\bibnamefont {Hofstaetter}}, \ and\ \bibinfo
  {author} {\bibfnamefont {B.}~\bibnamefont {Feigelson}},\ }\href {\doibase
  10.1103/PhysRevB.60.5392} {\bibfield  {journal} {\bibinfo  {journal} {Phys.
  Rev. B}\ }\textbf {\bibinfo {volume} {60}},\ \bibinfo {pages} {5392}
  (\bibinfo {year} {1999})}\BibitemShut {NoStop}%
\bibitem [{\citenamefont {Stoll}\ and\ \citenamefont
  {Schweiger}(2006)}]{Stoll2006}%
  \BibitemOpen
  \bibfield  {author} {\bibinfo {author} {\bibfnamefont {S.}~\bibnamefont
  {Stoll}}\ and\ \bibinfo {author} {\bibfnamefont {A.}~\bibnamefont
  {Schweiger}},\ }\href {\doibase 10.1016/j.jmr.2005.08.013} {\bibfield
  {journal} {\bibinfo  {journal} {J. Magn. Reson.}\ }\textbf {\bibinfo {volume}
  {178}},\ \bibinfo {pages} {42} (\bibinfo {year} {2006})}\BibitemShut
  {NoStop}%
\bibitem [{\citenamefont {Itoh}\ and\ \citenamefont
  {Watanabe}(2014)}]{Itoh2014}%
  \BibitemOpen
  \bibfield  {author} {\bibinfo {author} {\bibfnamefont {K.~M.}\ \bibnamefont
  {Itoh}}\ and\ \bibinfo {author} {\bibfnamefont {H.}~\bibnamefont
  {Watanabe}},\ }\href {\doibase 10.1557/mrc.2014.32} {\bibfield  {journal}
  {\bibinfo  {journal} {MRS Commun.}\ }\textbf {\bibinfo {volume} {4}},\
  \bibinfo {pages} {143} (\bibinfo {year} {2014})}\BibitemShut {NoStop}%
\bibitem [{\citenamefont {Rogers}\ \emph
  {et~al.}(2014{\natexlab{b}})\citenamefont {Rogers}, \citenamefont {Jahnke},
  \citenamefont {Teraji}, \citenamefont {Marseglia}, \citenamefont
  {M{\"{u}}ller}, \citenamefont {Naydenov}, \citenamefont {Schauffert},
  \citenamefont {Kranz}, \citenamefont {Isoya}, \citenamefont {McGuinness},\
  and\ \citenamefont {Jelezko}}]{Rogers2014b}%
  \BibitemOpen
  \bibfield  {author} {\bibinfo {author} {\bibfnamefont {L.~J.}\ \bibnamefont
  {Rogers}}, \bibinfo {author} {\bibfnamefont {K.~D.}\ \bibnamefont {Jahnke}},
  \bibinfo {author} {\bibfnamefont {T.}~\bibnamefont {Teraji}}, \bibinfo
  {author} {\bibfnamefont {L.}~\bibnamefont {Marseglia}}, \bibinfo {author}
  {\bibfnamefont {C.}~\bibnamefont {M{\"{u}}ller}}, \bibinfo {author}
  {\bibfnamefont {B.}~\bibnamefont {Naydenov}}, \bibinfo {author}
  {\bibfnamefont {H.}~\bibnamefont {Schauffert}}, \bibinfo {author}
  {\bibfnamefont {C.}~\bibnamefont {Kranz}}, \bibinfo {author} {\bibfnamefont
  {J.}~\bibnamefont {Isoya}}, \bibinfo {author} {\bibfnamefont {L.~P.}\
  \bibnamefont {McGuinness}}, \ and\ \bibinfo {author} {\bibfnamefont
  {F.}~\bibnamefont {Jelezko}},\ }\href {\doibase 10.1038/ncomms5739}
  {\bibfield  {journal} {\bibinfo  {journal} {Nat. Commun.}\ }\textbf {\bibinfo
  {volume} {5}},\ \bibinfo {pages} {4739} (\bibinfo {year}
  {2014}{\natexlab{b}})}\BibitemShut {NoStop}%
\bibitem [{\citenamefont {Edmonds}\ \emph {et~al.}(2012)\citenamefont
  {Edmonds}, \citenamefont {D'Haenens-Johansson}, \citenamefont {Cruddace},
  \citenamefont {Newton}, \citenamefont {Fu}, \citenamefont {Santori},
  \citenamefont {Beausoleil}, \citenamefont {Twitchen},\ and\ \citenamefont
  {Markham}}]{Edmonds2012}%
  \BibitemOpen
  \bibfield  {author} {\bibinfo {author} {\bibfnamefont {A.~M.}\ \bibnamefont
  {Edmonds}}, \bibinfo {author} {\bibfnamefont {U.~F.~S.}\ \bibnamefont
  {D'Haenens-Johansson}}, \bibinfo {author} {\bibfnamefont {R.~J.}\
  \bibnamefont {Cruddace}}, \bibinfo {author} {\bibfnamefont {M.~E.}\
  \bibnamefont {Newton}}, \bibinfo {author} {\bibfnamefont {K.-M.~C.}\
  \bibnamefont {Fu}}, \bibinfo {author} {\bibfnamefont {C.}~\bibnamefont
  {Santori}}, \bibinfo {author} {\bibfnamefont {R.~G.}\ \bibnamefont
  {Beausoleil}}, \bibinfo {author} {\bibfnamefont {D.~J.}\ \bibnamefont
  {Twitchen}}, \ and\ \bibinfo {author} {\bibfnamefont {M.~L.}\ \bibnamefont
  {Markham}},\ }\href {\doibase 10.1103/PhysRevB.86.035201} {\bibfield
  {journal} {\bibinfo  {journal} {Phys. Rev. B}\ }\textbf {\bibinfo {volume}
  {86}},\ \bibinfo {pages} {035201} (\bibinfo {year} {2012})}\BibitemShut
  {NoStop}%
\bibitem [{\citenamefont {Goss}\ and\ \citenamefont
  {Briddon}(2006)}]{Goss2006}%
  \BibitemOpen
  \bibfield  {author} {\bibinfo {author} {\bibfnamefont {J.}~\bibnamefont
  {Goss}}\ and\ \bibinfo {author} {\bibfnamefont {P.}~\bibnamefont {Briddon}},\
  }\href {\doibase 10.1103/PhysRevB.73.085204} {\bibfield  {journal} {\bibinfo
  {journal} {Phys. Rev. B}\ }\textbf {\bibinfo {volume} {73}},\ \bibinfo
  {pages} {085204} (\bibinfo {year} {2006})}\BibitemShut {NoStop}%
\bibitem [{\citenamefont {Mainwood}(1994)}]{Mainwood1994}%
  \BibitemOpen
  \bibfield  {author} {\bibinfo {author} {\bibfnamefont {A.}~\bibnamefont
  {Mainwood}},\ }\href {\doibase 10.1103/PhysRevB.49.7934} {\bibfield
  {journal} {\bibinfo  {journal} {Phys. Rev. B}\ }\textbf {\bibinfo {volume}
  {49}},\ \bibinfo {pages} {7934} (\bibinfo {year} {1994})}\BibitemShut
  {NoStop}%
\bibitem [{\citenamefont {Pinto}\ \emph {et~al.}(2012)\citenamefont {Pinto},
  \citenamefont {Jones}, \citenamefont {Palmer}, \citenamefont {Goss},
  \citenamefont {Briddon},\ and\ \citenamefont {{\"{O}}berg}}]{Pinto2012}%
  \BibitemOpen
  \bibfield  {author} {\bibinfo {author} {\bibfnamefont {H.}~\bibnamefont
  {Pinto}}, \bibinfo {author} {\bibfnamefont {R.}~\bibnamefont {Jones}},
  \bibinfo {author} {\bibfnamefont {D.~W.}\ \bibnamefont {Palmer}}, \bibinfo
  {author} {\bibfnamefont {J.~P.}\ \bibnamefont {Goss}}, \bibinfo {author}
  {\bibfnamefont {P.~R.}\ \bibnamefont {Briddon}}, \ and\ \bibinfo {author}
  {\bibfnamefont {S.}~\bibnamefont {{\"{O}}berg}},\ }\href {\doibase
  10.1002/pssa.201200050} {\bibfield  {journal} {\bibinfo  {journal} {Phys.
  status solidi}\ }\textbf {\bibinfo {volume} {209}},\ \bibinfo {pages} {1765}
  (\bibinfo {year} {2012})}\BibitemShut {NoStop}%
\bibitem [{\citenamefont {Wassell}\ \emph {et~al.}(2018)\citenamefont
  {Wassell}, \citenamefont {McGuinness}, \citenamefont {Hodges}, \citenamefont
  {Lanigan}, \citenamefont {Fisher}, \citenamefont {Martineau}, \citenamefont
  {Newton},\ and\ \citenamefont {Lynch}}]{Wassell2018}%
  \BibitemOpen
  \bibfield  {author} {\bibinfo {author} {\bibfnamefont {A.~M.}\ \bibnamefont
  {Wassell}}, \bibinfo {author} {\bibfnamefont {C.~D.}\ \bibnamefont
  {McGuinness}}, \bibinfo {author} {\bibfnamefont {C.}~\bibnamefont {Hodges}},
  \bibinfo {author} {\bibfnamefont {P.~M.}\ \bibnamefont {Lanigan}}, \bibinfo
  {author} {\bibfnamefont {D.}~\bibnamefont {Fisher}}, \bibinfo {author}
  {\bibfnamefont {P.~M.}\ \bibnamefont {Martineau}}, \bibinfo {author}
  {\bibfnamefont {M.~E.}\ \bibnamefont {Newton}}, \ and\ \bibinfo {author}
  {\bibfnamefont {S.~A.}\ \bibnamefont {Lynch}},\ }\href {\doibase
  10.1002/pssa.201800292} {\bibfield  {journal} {\bibinfo  {journal} {Phys.
  Status Solidi Appl. Mater. Sci.}\ }\textbf {\bibinfo {volume} {215}}
  (\bibinfo {year} {2018}),\ 10.1002/pssa.201800292}\BibitemShut {NoStop}%
\bibitem [{\citenamefont {Dhomkar}\ \emph {et~al.}(2018)\citenamefont
  {Dhomkar}, \citenamefont {Zangara}, \citenamefont {Henshaw},\ and\
  \citenamefont {Meriles}}]{Dhomkar2018}%
  \BibitemOpen
  \bibfield  {author} {\bibinfo {author} {\bibfnamefont {S.}~\bibnamefont
  {Dhomkar}}, \bibinfo {author} {\bibfnamefont {P.~R.}\ \bibnamefont
  {Zangara}}, \bibinfo {author} {\bibfnamefont {J.}~\bibnamefont {Henshaw}}, \
  and\ \bibinfo {author} {\bibfnamefont {C.~A.}\ \bibnamefont {Meriles}},\
  }\href {\doibase 10.1103/PhysRevLett.120.117401} {\bibfield  {journal}
  {\bibinfo  {journal} {Phys. Rev. Lett.}\ }\textbf {\bibinfo {volume} {120}},\
  \bibinfo {pages} {117401} (\bibinfo {year} {2018})}\BibitemShut {NoStop}%
\end{thebibliography}
\end{document}